\documentclass{sig-alternate-05-2015}

\usepackage{booktabs} 
\usepackage{xspace}
\usepackage[	n,
	operators,
	advantage,
	sets,
	adversary,
	landau,
	probability,
	notions,	
	logic,
	ff,
	mm,
	primitives,
	events,
	complexity,
	asymptotics,
	keys]{cryptocode}
\usepackage{dashbox}
\usepackage{acronym}
\usepackage{enumitem}
\usepackage{float}
\usepackage[skip=4pt,font=small]{caption}

\usepackage{subfig}
\usepackage{graphicx}

\begin{document}

\setcopyright{acmcopyright}





%

\title{Scalable Attestation Resilient to Physical Attacks\\for Embedded Devices in Mesh Networks}




%
%
%
%
%

\numberofauthors{2}
\author{
\alignauthor
Florian Kohnh\"auser\\
       \affaddr{Technische Unversitat Darmstadt, Germany}\\
       \email{\normalsize kohnhaeuser@seceng.informatik.tu-darmstadt.de}
\alignauthor
Niklas B\"uscher\\
       \affaddr{Technische Unversitat Darmstadt, Germany}\\
       \email{\normalsize buescher@seceng.informatik.tu-darmstadt.de}
\and
\alignauthor
Sebastian Gabmeyer\\
       \affaddr{Technische Unversitat Darmstadt, Germany}\\
       \email{\normalsize gabmeyer@seceng.informatik.tu-darmstadt.de}
\alignauthor
Stefan Katzenbeisser\\
       \affaddr{Technische Unversitat Darmstadt, Germany}\\
       \email{\normalsize katzenbeisser@seceng.informatik.tu-darmstadt.de}
}

\maketitle

\begin{abstract}
Interconnected embedded devices are increasingly used in various scenarios, including industrial control, building automation, or emergency communication.~As these systems commonly process sensitive information or perform safety critical tasks, they become appealing targets for cyber attacks.~A promising technique to remotely verify the safe and secure operation of networked embedded devices is remote attestation.~However, existing attestation protocols only protect against software attacks or show very limited scalability. In this paper, we present the first scalable attestation protocol for interconnected embedded devices that is resilient to physical attacks. Based on the assumption that physical attacks require an adversary to capture and disable devices for some time, our protocol identifies devices with compromised hardware and software. Compared to existing solutions, our protocol reduces communication complexity and runtimes by orders of magnitude, precisely identifies compromised devices, supports highly dynamic and partitioned network topologies, and is robust against failures. We show the security of our protocol and evaluate it in static as well as dynamic network topologies. Our results demonstrate that our protocol is highly efficient in well-connected networks and robust to network disruptions.
\end{abstract}


\newtheorem{definition}{Definition}
\newtheorem{theorem}{Theorem}

\newcommand{\ourname}{SCAP\xspace}

\newcommand{\para}[1]{\smallskip\noindent\textbf{#1.}}


\newcommand{\AEnc}[2]{\ensuremath{\textsf{AEnc}(#1, #2)}}
\newcommand{\ADec}[2]{\ensuremath{\textsf{ADec}(#1, #2)}}
\newcommand{\ADecAbort}[2]{\ensuremath{\textsf{ADecOrAbort}(#1, #2)}}
\newcommand{\nonce}{\textrm{nonce}}
\newcommand{\sessionkey}{\ensuremath{\key_{ses}}}
\newcommand{\devicekey}{\ensuremath{\textsf{dk}}}
\newcommand{\publickey}{\ensuremath{\textsf{pk}}}
\newcommand{\privatekey}{\ensuremath{\textsf{sk}}}

\newcommand{\msg}{\textrm{msg}}

\newcommand{\clock}{\textrm{RROC()}\xspace}
\newcommand{\Tslot}{\ensuremath{\delta}}
\newcommand{\now}{\ensuremath{t}}
\newcommand{\timecheck}{\textsf{Checktime}}
\newcommand{\timecheckHB}[1]{\textsf{Checktime(#1)} = \HB}
\newcommand{\timecheckLE}[1]{\textsf{Checktime(#1)} = \LE}
\newcommand{\hb}{\textsf{hb}}
\newcommand{\HB}{\textsf{HB}} 
\newcommand{\LE}{\textsf{LE}} 
\newcommand{\curHB}{\ensuremath{\hb_{cur}}}
\newcommand{\nextHB}{\ensuremath{\hb_{next}}}
\newcommand{\tattack}{\ensuremath{t_{attack}}\xspace}

\newcommand{\agg}[1]{\ensuremath{\textsf{agg(}#1\textsf{)}}}
\newcommand{\merge}[1]{\ensuremath{\textsf{merge(}#1\textsf{)}}}

\newcommand{\Adv}{\textsf{Adv}\xspace}
\newcommand{\Net}{\textsf{Net}\xspace}
\newcommand{\Exp}{\textsf{SECATT}\xspace}
\newcommand{\ExpA}{\Exp^{n,c}_{\Adv}\xspace}
\newcommand{\operator}{\ensuremath{\mathcal{O}\xspace}}

\newcommand{\device}[1]{\ensuremath{\mathcal{D}_{#1}}\xspace}
\newcommand{\generated}{\ensuremath{\textrm{flag}_{gen}}\xspace}
\newcommand{\leader}{\ensuremath{\textrm{flag}_{leader}}\xspace}
\newcommand{\TEE}{\textrm{Execute in TEE:}}
\newcommand{\exitTEE}{\pcreturn}

\acrodef{TEE}{Trusted Execution Environment}
\acrodef{ROM}{Read-Only Memory}
\acrodef{MPU}{Memory Protection Unit}
\acrodef{DoS}{Denial of Service}


\newcommand{\todoF}[1]{\textcolor{red}{\,\textbf{ToDo Florian:} #1\xspace}\,}
\newcommand{\todoN}[1]{\textcolor{purple}{\,\textbf{ToDo Niklas:} #1\xspace}\,}

\renewcommand{\textfraction}{0.01} 
\renewcommand{\floatpagefraction}{0.95} 
\renewcommand{\topfraction}{0.8}
\renewcommand{\bottomfraction}{0.8} 
\setlength{\emergencystretch}{1em}
\setlength{\belowcaptionskip}{-6pt}

\section{Introduction}
\label{intro}

Nowadays, networked embedded devices are increasingly present in every aspect of our lives. This paradigm, often referred to as the Internet of Things (IoT), is expected to constantly evolve in scale and complexity, reaching 20.8 billion devices by 2020~\cite{gartner2015iot}. Technologies like Bluetooth Smart, IEEE 802.15.4, Wi-Fi Direct, ZigBee, or Z-Wave enable embedded devices to form large wireless mobile ad hoc networks (MANETs). In MANETs, all devices cooperate in the distribution of data in the network, thus establishing a decentralized and self-organized network topology. Interconnected embedded devices are frequently used in industrial control, building automation, military communication, or sensor networks. As such systems often process privacy-sensitive information or perform safety-critical tasks, their malfunction or misuse can cause serious damage. Unfortunately, software for embedded systems is typically written in unsafe programming languages and often reluctantly maintained. Additionally, even though an adversary requires significant resources to physically tamper with a device~\cite{becher2006tampering}, (secure) hardware on embedded systems is usually not hardened against physical tampering; thus, interconnected embedded devices are appealing targets for cyber attacks~\cite{morgner2016all, park2016ain, slawomir2016gattacking}.


To detect and mitigate such attacks, it is important to monitor the correct operation of embedded devices and detect any malfunctioning or misuse as early as possible. For this purpose, attestation protocols have been introduced, which allow a third party, the \emph{verifier}, to check the integrity of a remote device, the \emph{prover}. Since traditional single device attestation protocols are impractical in large mesh networks due to their overhead of attesting each device individually, scalable attestation protocols have recently been proposed~\cite{ambrosin2016sana, asokan2015seda}. These protocols perform an efficient attestation of large networks by distributing the attestation burden across all devices in the network. All scalable attestation protocols are based on the assumption that an adversary can only manipulate the software of provers. Thus, they cannot withstand an adversary who is able to perform physical attacks and tamper with the hardware of provers. Yet, an adversary can rather easily capture a device and tamper with its hardware as devices forming MANETs are often distributed over wide public areas and consist of a multitude of devices.~Hence, a scalable attestation protocol that is resilient to physical attacks is much needed.

Ibrahim et al.~\cite{ibrahimdarpa} presented a first approach to solve this problem by combining existing scalable attestation approaches~\cite{ambrosin2016sana, asokan2015seda} with absent detection~\cite{conti2009mobility} to detect both software and hardware attacks. The absent detection protocol is based on the assumption that a strong adversary, who physically tampers with a device, must temporarily take the device offline for a certain amount of time, e.g., to disassemble the device and extract secret keys~\cite{becher2006tampering}. To detect offline and thus physically compromised devices, each device periodically emits a heartbeat that needs to be received, verified, and logged by every other device in the network. Although a functional solution to the problem, the protocol suffers from several shortcomings. First, the amount of exchanged messages per heartbeat period scales quadratically with the number of devices in the network. This causes scalability issues in large networks with respect to network communication, energy consumption, and runtime performance. Furthermore, the protocol is very error-prone, since a single defective transmission of a heartbeat suffices to cause a false positive, where a healthy device is mistakenly regarded as compromised. Aggravating this, the protocol is only able to attest the state of the overall network and cannot identify particular compromised devices. Hence, a single false positive causes the entire network to be considered as compromised. Finally, the protocol relies on the assumption that during protocol execution the network topology is static and connected, which is a very strong limitation for wireless mesh networks.

In this paper, we present the first scalable attestation protocol (\ourname) for interconnected embedded devices that is resilient to physical attacks. To protect against strong adversaries, we build on the established assumption that an adversary needs to take a device offline to physically tamper with it~\cite{becher2006tampering, conti2009mobility, ibrahimdarpa}. In our protocol, a single leader device periodically emits a new heartbeat that is propagated in the network. To obtain the newest heartbeat from a neighboring device, a device must authenticate itself with the previous heartbeat. Since a device that is under physical attack has to be absent for at least one heartbeat period, it will miss this period's heartbeat and thus be unable to obtain any further heartbeats. To prevent a collusion between compromised devices, heartbeats are stored in lightweight secure hardware and transmitted encrypted via secure channels. During the actual attestation, devices that fail to authenticate with the newest heartbeat are regarded as physically compromised, whereas devices with a compromised software are detected based on existing software attestation techniques. In case of an outage of the leader, a new leader device is determined through a leader election process. By optionally storing the attestation result in each device, our protocol is able to efficiently attest highly dynamic and partitioned network topologies.

We show that our protocol is secure against an adversary who compromises all but one device in the network.~Finally, we demonstrate the practicability of our protocol in static and dynamic networks.~In summary, \ourname provides the following improvements over existing work:

%


\begin{itemize}
\setlength{\itemsep}{-1pt}

\item \ourname can precisely identify devices whose hardware and/or software is compromised, if less than half of all devices in the network are compromised.

\item \ourname is very efficient. Compared to the best previous work~\cite{ibrahimdarpa}, we reduce the number of sent messages per time period from $O(n^2)$ to $O(n)$\footnote{\small In fact, when detecting physically compromised devices through their absence, $O(n)$ transmitted messages per time period is the best possible solution, since each device must at least send or receive one message to show that it is present.}, thus, achieving scalability to millions of devices (where $n$ denotes the total number of devices in the network).
\item \ourname is robust against network and device failures by\\ (1) relying on a one-to-many delay-tolerant link in contrast to a many-to-many continuous link, as used in the best previous work~\cite{ibrahimdarpa}, and\\ (2) offering a recovery mechanism, the leader election protocol, that minimizes the amount of false negatives.

\item \ourname provides a novel efficient aggregation scheme, e.g., attests of $4,000$ devices fit into 1kB. This allows to attest highly dynamic and partitioned network topologies efficiently.

\item \ourname is the first scalable attestation protocol that is evaluated in dynamic network topologies.


\end{itemize}



\para{Outline} The rest of the paper is organized as follows. In \S~\ref{relatedwork} we summarize existing work. In \S~\ref{preliminaries} the system model, device requirements, and adversary model are presented. In \S~\ref{protocol}, we describe our novel attestation approach to detect physically compromised devices. Then, in \S~\ref{extensions} we extend the attestation protocol to execute a recovery protocol on failures, verify the software integrity of devices, and support dynamic topologies during attestation. The performance of \ourname is evaluated in \S~\ref{evaluation}. Finally, we conclude in \S~\ref{conclusion}.


\section{Related Work}
\label{relatedwork}

\para{Device Attestation}
Remote attestation is a mechanism that allows a third party, the
verifier, to check the integrity of a remote system, the
prover. Protocols that target the attestation of a single embedded
device are either
software-based~\cite{kovah2012new,li2011viper} or
hardware-based~\cite{brasser2015tytan, eldefrawy2012smart, noorman2013sancus}. Software-based techniques
require no secure hardware, but rely on assumptions that have been
shown to be hard to achieve in
practice~\cite{armknecht2013security}. Hardware-based attestation
mechanisms provide much stronger security guarantees by relying on
lightweight security architectures. Nevertheless, single-device
approaches are impractical in mesh networks due to the large overhead
of attesting each device individually.

Recently, protocols started to focus on an efficient attestation of
multiple embedded devices. Park et al.~\cite{park2012smatt} proposed
to compare the integrity measurements of multiple devices. Yet, their
approach requires identical devices and only enables a probabilistic
attack detection rate. Asokan et al.~\cite{asokan2015seda} present a
highly efficient attestation scheme for large-scale networks of
embedded devices that requires only \ac{ROM} and a simple \ac{MPU}. In
their scheme, each device attests its neighbors and reports the
aggregated result back to its parent, eventually received by the
verifier. Ambrosin et al.~\cite{ambrosin2016sana} enhance this work
by introducing a novel signature scheme that enables anyone to
publicly verify the attestation result and allows the network to
contain untrustworthy aggregator devices, such as routers or cloud
severs. Yet, besides the work by Ibrahim et. al~\cite{ibrahimdarpa}, which has been discussed in \S~\ref{intro}, existing works consider the adversary to compromise only the software on devices. In mesh networks, this assumption may not hold, since an adversary can comparatively easy capture a device and physically tamper with it.

\para{Capture Detection}
Several works have been proposed on the detection of node capture attacks, where an adversary physically approaches and manipulates a device. They all build on the assumption that an adversary needs to take a device offline, in order to tamper with it~\cite{becher2006tampering}. Conti et al. suggested that a node is collaboratively flagged as captured if it fails to re-meet with any other node within a fixed time interval~\cite{conti2009mobility, conti2010smallville}. In the approach by Ho~\cite{ho2010distributed}, nodes use statistical methods to detect absent neighbor devices in static network topologies. Recently, Agrwal et al. proposed to deploy multiple TPM-equipped cluster heads in the network, which check the integrity of the software as well as the physical presence of all nodes in the cluster~\cite{agrwal15:programIntegrity}. Nevertheless, existing approaches are unable to detect devices with compromised software~\cite{conti2009mobility, conti2010smallville,ho2010distributed}, require the deployment of additional hardware~\cite{agrwal15:programIntegrity}, are only applicable in static network topologies~\cite{ho2010distributed}, or lack scalability~\cite{ibrahimdarpa}.






\para{Secure Data Aggregation} Since ad hoc networks are often deployed to collect sensory data, many efficient and integrity-preserving aggregation schemes for mesh networks have been proposed. Unfortunately, these schemes rely on very costly asymmetric cryptographic operations~\cite{castelluccia09:homomorphicAggregation, przydatek03:sia},
require to maintain a specific network topology during aggregation~\cite{hu03:secureAggregationWSN,przydatek03:sia}, or need multiple communication rounds~\cite{taban2008efficient}, which both is undesirable, as
it leads to communication overhead in dynamic network topologies. Thus, a lightweight aggregation scheme suitable for remote attestation of embedded devices that supports dynamic topologies and allows the identification of compromised devices is missing.

\section{Preliminaries}
\label{preliminaries}

\para{System Model}
In our model, we consider embedded devices that can be heterogeneous in terms of hardware capabilities and software resources, e.g., devices with different software, computational power, storage capacity, or security functionalities. All embedded devices are connected in a mesh network topology. This topology can be static, where devices remain stationary and the network is connected, or dynamic, where devices can move freely and the network can be temporarily partitioned. However, in dynamic network topologies, we assume that devices meet each other regularly due to their mobility. Devices that are unreachable for some time $\Tslot$ are regarded as compromised, since it is uncertain whether they will ever contribute to the network again. We further assume that each device \device{i} gets initialized and deployed by a trusted \emph{network operator} $\mathcal{O}$, once (\S~\ref{sec:protocol:initialization}).

After deployment, the goal of $\mathcal{O}$ is to ensure the correct and safe operation of all devices $\device{1}, \device{2}, ..., \device{n}$ in the network. Therefore, $\mathcal{O}$ regularly verifies the integrity of all devices by executing the proposed attestation protocol. The attestation protocol determines all devices whose software is in a trustworthy, i.e., unmanipulated and up-to-date, state and whose hardware has not been tampered with. We refer to these devices as \emph{healthy} devices, in contrast to \emph{compromised} devices. Executing the protocol, $\mathcal{O}$ is able to learn the precise identity of all healthy and all compromised devices. This may serve as a first step towards physically locating and recovering compromised devices. In order to perform the attestation protocol, $\mathcal{O}$ requires a connection to at least one device in the network.


\para{Device Requirements}
\label{pre:req:dev}
We assume that each device \device{i} provides the minimal hardware properties for remote attestation, according to the work by Francillon et al.~\cite{francillon2014minimalist}. In practice, these properties can be implemented with \ac{ROM} and a simple \ac{MPU}. \ac{ROM} stores the protocol code and cryptographic keys, and the \ac{MPU} ensures an uninterruptible execution of the protocol code and allows only protocol code to access the cryptographic keys. Recently, it has been shown that these minimal hardware properties are available even on many low-cost commodity embedded devices~\cite{kohnhauser2016secure}. Additionally, our attestation protocol relies on authentic time measurements. In order to prevent malware from tampering with the device clock, each device must provide a write-protected real-time clock. Protected real-time clocks are already built-in many existing commodity embedded devices~\cite{stmrtc, mspusersguide}. We henceforth refer to the execution space, where all required hardware properties are fulfilled, as \acf{TEE}.


\bigskip
\para{Adversary Model}
\label{pre:adv}
In this work, we regard a powerful adversary \Adv, who is able to
mount attacks on the network as well as the software and hardware of devices.
In detail, \Adv is granted \emph{full control} over all
messages in the network (Dolev-Yao model).
Thus, \Adv can eavesdrop, modify,
delete, or synthesize all message between any two entities.\footnote{\small We note that the model allows \ac{DoS} attacks, such as jamming or cutting wires. These attacks cannot be prevented against a physically present adversary. However, \ac{DoS} attacks have no influence on the security of our scheme, as \Adv cannot use them to forge a healthy system state.}
Moreover, \Adv is allowed to compromise the software of \emph{all}
devices in the network. This gives \Adv full control over the devices'
execution state and storage, yet, no access to the protected contents
inside the \ac{TEE}.
We further allow \Adv to capture and physically tamper with up to \emph{all 
but one} device in the network, when attesting the overall network state,
and up to \emph{half of all} devices in the network, when knowledge on the
precise identity of compromised devices is required. 
For the physically compromised devices,
\Adv is able to access device secrets and code inside the \ac{TEE}
and is allowed to manipulate the clock. We note that it is impossible
to guarantee a secure device attestation, if all devices in the network
have physically been compromised~\cite{ibrahimdarpa}.
Finally, as in~\cite{ibrahimdarpa} we assume that mounting a physical attack
requires at least a time $\tattack$, in which the device is offline,
e.g., to decapsulate the device and to launch a microprobing attack.
Depending on the device's level of tamper resistance and the
adversaries resources, such attacks typically require hours up to
weeks in specialized laboratory environments~\cite{skorobogatov2012physical}.

\section{\ourname}
\label{protocol}
In the following, we describe the \ourname protocol, which identifies devices in the network have physically been tampered with. Note that the detection of hybrid attacks, i.e., attacks that target hardware and software, is discussed in the next section (\S~\ref{extensions:software}).  \ourname consists of three different phases. In the \textit{initialization phase} (\S~\ref{sec:protocol:initialization}), the trusted network operator $\mathcal{O}$ initializes each device once, before the deployment of the network. The \textit{heartbeat phase} (\S~\ref{sec:protocol:heartbeat}) is periodically executed during the operation of the network. In this phase, all physically uncompromised devices maintain a valid state by sharing a common group key, namely the heartbeat. We will show how the heartbeat is periodically regenerated and propagated in the network and demonstrate that physically compromised devices are unable to obtain the heartbeat. Finally, in the \textit{attestation phase} (\S~\ref{sec:protocol:attestation}), $\mathcal{O}$ initiates an attestation of the network and obtains a report, which exhibits all physically compromised devices.

\begin{figure*}[t]
\input{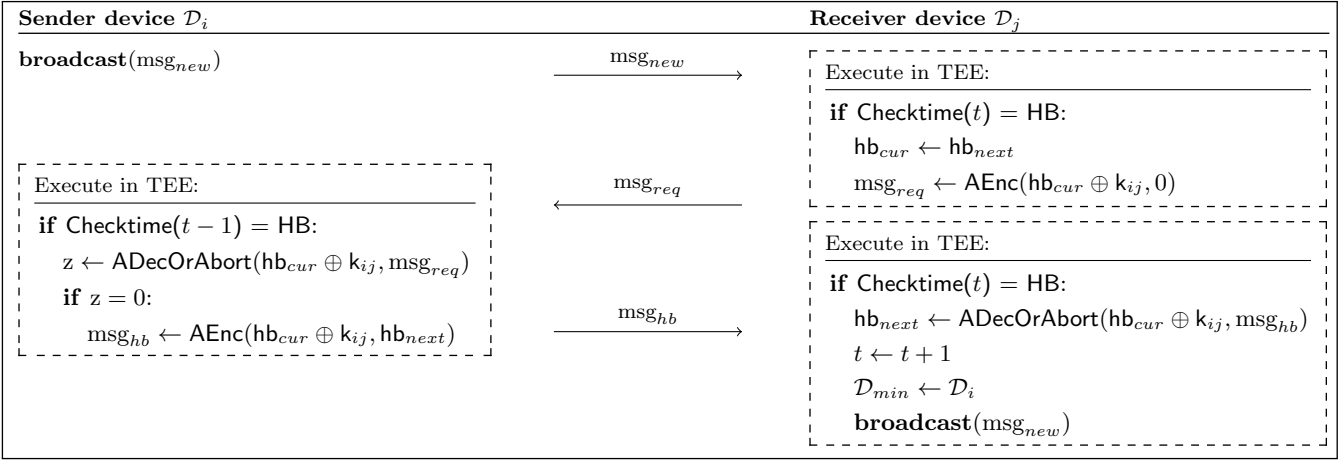}
\caption{The heartbeat transmission protocol between a sender $\device{i}$ and receiver $\device{j}$ after secure channel establishment, i.e, both devices share a channel key $k_{ij}$ and know their identities.\label{proto:heartbeat}}
\end{figure*}

\subsection{Initialization Phase}
\label{sec:protocol:initialization}
\para{Preliminaries}
Devices can either be in a \textit{healthy} or \textit{compromised} hardware
state. We discretize the time into non-overlapping time periods
$t \in \{1,2,3,...\}$ of fixed length $\Tslot$. We reference the starting times
of each time period with $T_1, T_2, T_3, ...$. The real time $T_{clock}$ can be
read by any device from a reliable read only clock \clock, which for simplicity
is assumed to be synchronized between all devices. Each devices keeps track of
the current time period $t$, running from time $T_t$ until $T_{t+1}$.  In the
remainder of this section, we assume an implementation of a function
$\timecheck(t)$  that returns a constant \HB, iff the real time is within the
time period indicated by parameter $t$, i.e, $T_t \leq T_{clock} < T_{t+1}$
and otherwise $\false$. 

\begin{figure*}[t]
\input{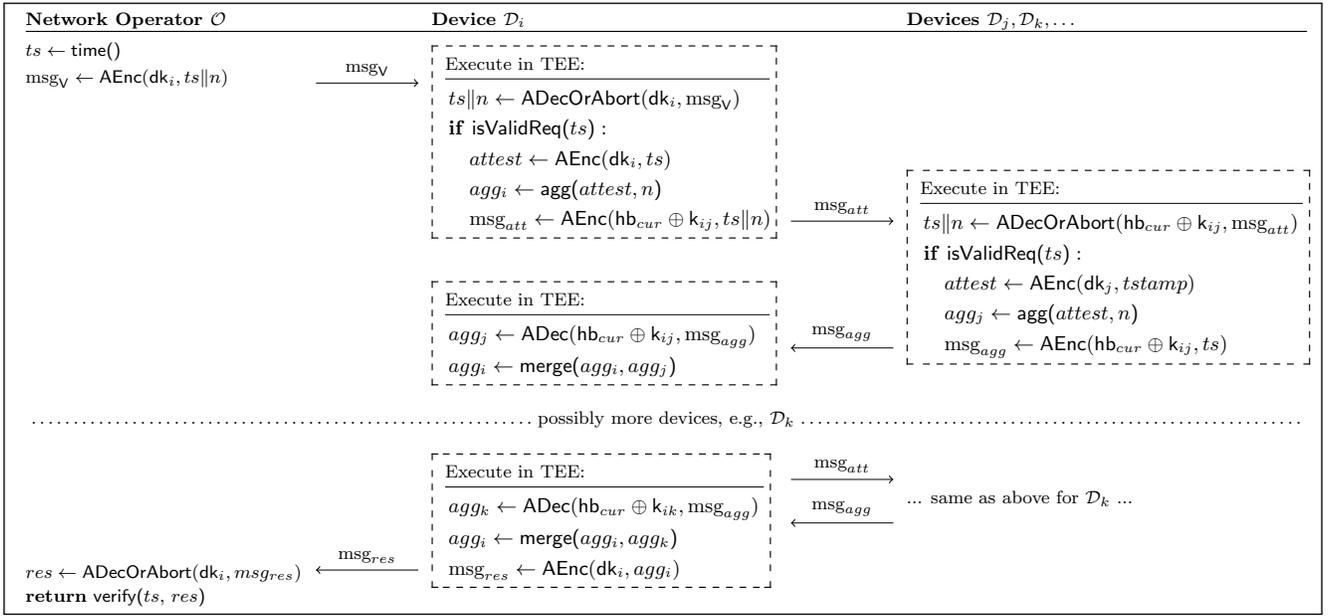}
\caption{The attestation protocol after secure channel establishment, i.e, all devices share a channel key $k$ and know their identities. \label{protocol:attestation}}
\end{figure*}

\para{Enrollment}
In the enrollment phase, the network operator $\mathcal{O}$ initializes the
\ac{TEE} of all devices with the following secrets. First, devices store two 
initial heartbeats $\curHB$ and $\nextHB$, which function as a group secret between
all healthy devices.
Second, each device is equipped with a device-dependent symmetric key $\devicekey_i$,
used during attestation to generate a device unique attest,
and an asymmetric key pair ($\pk_i, \sk_i$), employed to establish
secure channels between devices. Finally, devices record the current time period
$t$, their own device identifier \device{i}, and the identifier of the leader device $\device{min}$, which is the
first device \device{1} in the network. Table~\ref{tab:protocol:acro} provides
a summary of relevant definitions.

For explanatory reasons, we assume an initial enrollment of all devices. However, \ourname also allows devices to be enrolled at any point in time by issuing the current heartbeat.


\begin{table}[h]
\begin{center}
\begin{tabular}{ll}
\textit{Acronym} & \textit{Usage} \\ 
\hline 
$\Tslot$ & length of heartbeat period\\
$t$ & current time period  \\ 
$\device{i}$ & unique device identifier\\
$\device{min}$ & device identifier of the leader device\\
$\curHB$ & current valid heartbeat\\
$\nextHB$ & heartbeat valid in next time period\\
$\pk_i, \sk_i$ & key pair for channel establishment\\
$\key_{ij}, \key_{ik}, \dots$ & channel keys with neighbors $\device{j},\device{k},\dots$\\
$\devicekey_i$ & device key for attestation with operator
\end{tabular} 
\end{center}
\caption{Overview of all secrets stored in the $\device{i}$'s \ac{TEE}.\label{tab:protocol:acro}}
\end{table}

\begin{figure*}[t]
\subfloat[Heartbeat protocol.\label{fig:protocol:heartbeat}]{
  \includegraphics[width=.47\linewidth]{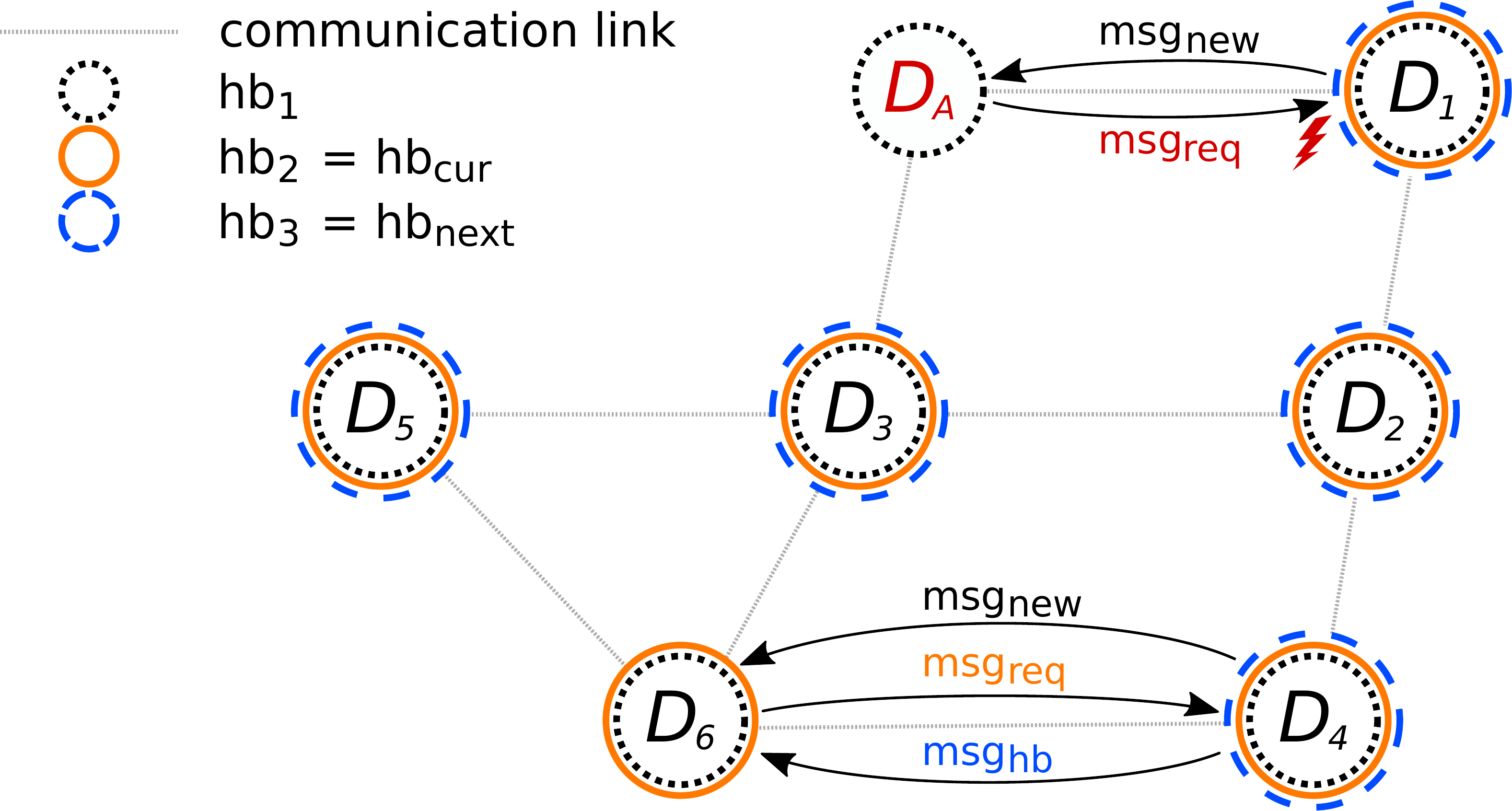}
}\hfill
\subfloat[Attestation protocol.\label{fig:protocol:attestation}]{
  \includegraphics[width=.47\linewidth]{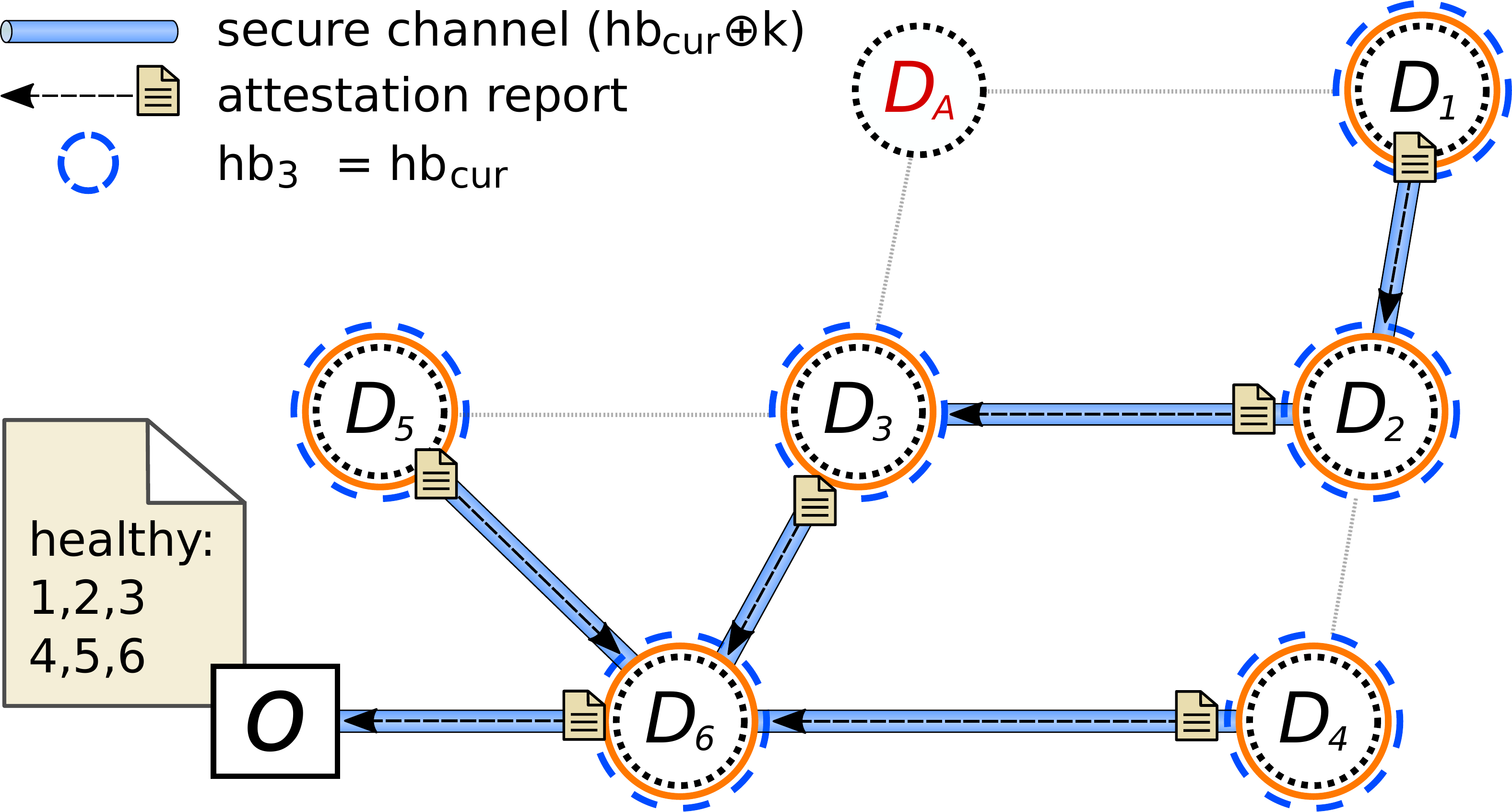}
}
\caption{In Fig~\ref{fig:protocol:heartbeat} the heartbeat protocol is illustrated for 7 devices in time period $t=2$. All devices 
store the initial heartbeat $\hb_{1}$ that was used to secure the exchange of $\hb_{2}$. Subsequently, $\hb_{2}$ is used to exchange
the next heartbeat $\hb_{3}$ for the upcoming time period $t=3$. Such an exchange is illustrated between $\device{4}$ and $\device{6}$. We observe that \device{A} was physically compromised in time period $t = 2$ and thus did not receive $hb_{2}$. Hence, \device{A} is also unable to obtain $hb_{3}$ or any following heartbeat. In Fig~\ref{fig:protocol:attestation}, the same network is
illustrated in time period $t=3$, while answering an attestation request by $\mathcal{O}$. This request was forwarded to all devices that
were in possession of the current heartbeat $\hb_{3}$. Consequently, \device{A}'s attest is not included in the attestation report, as it is excluded from all communication in the attestation protocol. Using a spanning tree topology, attestation reports are propagated back to $\mathcal{O}$ and aggregated in each hop.
\label{}}
\end{figure*}

\subsection{Heartbeat Phase}
\label{sec:protocol:heartbeat}

\para{Basic Idea}
The heartbeat protocol is the core protocol of our approach. It excludes
devices from the network that are offline for more than one time period and, hence,
are assumed to be physically tampered with. During protocol execution,
a so-called \textit{leader device} emits a new secret group key, named
\textit{heartbeat}, that is propagated in the network. Obtaining this heartbeat
requires a device to authenticate with the heartbeat of the previous time period.
Therefore, devices that are offline in an arbitrary time period $T_a$ miss the
heartbeat that is propagated in $T_a$ and thus are unable to obtain a heartbeat
in any subsequent time period $T_{a+1}, T_{a+2}, ...$. Since any communication
between devices in all protocols is secured using the newest heartbeat as a key,
physically compromised devices are unable to participate any more.
In the following, we describe the heartbeat transmission protocol, formalized in
Figure~\ref{proto:heartbeat}, which is run between two neighboring devices to
transfer the heartbeat from one device to the other.


\vspace{-1mm}
\para{Heartbeat Transmission Protocol}
The emission of the new heartbeat in every time period is initialized
by the leader device. As soon as the leader
observes that the real time $T_{clock}$ has reached the start of a new
time period ($\timecheck(t)$ returns $\HB$), the leader first
updates the heartbeat of the current time period $\curHB$ to the most recently
exchanged heartbeat $\nextHB$.
We remark that heartbeats could also be indexed by the time period in which
they are active in, e.g., $\hb_{1}$,$\hb_{2}$, $\hb_{3}, \dots$. However,
as only two heartbeats are relevant for any device, only these two, i.e.,
the current and next heartbeat, are stored and referenced.
After updating the current heartbeat, the leader samples a new heartbeat $\nextHB$
for the subsequent period $t+1$ and increments its time pointer $t$ by one.
Consequently, the time period described by the
pointer is now ahead of the real time $T_t > T_{clock}$. 
A time pointer ahead of the real time indicates a device that it
is in possession of a heartbeat for the upcoming time period.
The leader initialization code is illustrated below.

\vspace{-1.5mm}
\hspace{20mm}
\pseudocode{
\dbox{\begin{subprocedure}\procedure{\TEE}{
  \textrm{\pcif \timecheckHB{$t$}:}\\ 
  \t \text{$\curHB \leftarrow \nextHB$}\label{line:leader:update}\\
  \t \text{$\nextHB \sample \bin^n$}\\
  \t \text{$\now \leftarrow \now + 1$} \\
  \t \text{{\bf broadcast}($\msg_{new}$)}
}
\end{subprocedure}}
}


\noindent Next, the leader informs its
neighbors about the new heartbeat with a message $\msg_{new}$.
For simplicity, we henceforth assume that two neighboring devices
have already established a shared secret $\key_{ij}$ by performing
a key exchange using their public keys authenticated with the current heartbeat.

On receiving
$\msg_{new}$ from any device $\device{i}$, a device $\device{j}$ will enter
its \ac{TEE} and check whether the next time period has been reached.
If this is the case, $\device{j}$ will update
its current heartbeat to the previously communicated one. Afterwards,
$\device{j}$ encrypts a fixed string, e.g., '0', under the current
heartbeat $\curHB$ XOR-ed with the channel key $\key_{ij}$ shared by
both devices and sends the result to $\device{i}$. We refer to this
XOR-ed key, as the session key.
A healthy $\device{i}$ can decrypt the message by also computing the session
key. A successful decryption proves that $\device{j}$ is in possession of
the current heartbeat (and the channel key) and is therefore eligible for the 
next heartbeat. Then, $\device{i}$ answers with a
message $\msg_{hb}$ containing the next heartbeat $\nextHB$, also encrypted
with the session key. On successful decryption, device $\device{j}$ stores the
new heartbeat as $\nextHB$. Afterwards $\device{j}$ increments its time
period pointer and then announces this new heartbeat to its neighbors with
$\msg_{new}$.  Figure~\ref{fig:protocol:heartbeat} illustrates the heartbeat
transmission phase in a network with $6$ healthy devices and one adversarial
device \device{A} that was physically compromised in time period $t = 2$.

We note, that the heartbeat protocol relies on the availability of the leader
device, which constitutes a single point of failure. In
\S~\ref{extensions:leaderelection} we present an extension that makes 
the heartbeat protocol more robust against device outages, network 
partitioning, or targeted denial of service attacks.

\subsection{Attestation Phase}
\label{sec:protocol:attestation}

\smallskip\noindent\textbf{Basic idea.} The attestation protocol allows the operator
$\mathcal{O}$ to check the state of all devices in the network. For this purpose,
$\mathcal{O}$ issues an attestation request that is answered by all devices with
an attestation report. Propagating the attestation request through the network arranges a spanning tree whose root is $\mathcal{O}$. This enables an efficient transmission and aggregation of attestation reports along the spanning tree to $\mathcal{O}$.
\ourname supports two variants of attestation. The first variant allows to attest
the overall network state and is secure against an adversary who compromises all but one device.
However, it only outputs a Boolean result, namely whether all devices are healthy or not.
The second variant precisely identifies compromised devices by id and in this way
increases the protocol's robustness and applicability in practice. Yet, it requires
more than half of all devices in the network to be healthy.


\smallskip\noindent\textbf{Attestation protocol.} The protocol is formalized in
Figure~\ref{protocol:attestation}. The operator
$\mathcal{O}$ initially connects to a device \device{i} in the network and emits
an attestation request. The request contains the concatenation of a current
timestamp $ts$ and the number of devices $n$ in the network,
encrypted under the device's key $\devicekey_i$,
which is only shared between $\device{i}$ and $\mathcal{O}$.
By verifying the authenticity and timeliness of the request
($\textsf{isValidReq(}ts\textsf{)}$),
denial of service attacks through replays can be prevented. 
Next, the attestation request, consisting of the concatenation of $ts$ and $n$,
is propagated by \device{i} to its neighboring devices. This and all following
communication between two devices is secured with the pairwise session key,
i.e., the current heartbeat XOR-ed with the channel key. Any
device that receives an attestation request first verifies the request and
then also propagates the request to its neighboring devices. These
steps are repeated until the attestation request reaches devices, whose
neighbors already have received the request. In this way, a spanning tree is
constructed. Leaf devices that cannot propagate the request any further
return an attestation report to their parent device from which they initially obtained the attestation request.
The attestation report contains their own attest, which consists of $ts$ encrypted under their own device key.
Every non leaf device merges its own attest (and identifier) with all received attestation 
reports and propagates the merged report to its parent device.
Eventually, \device{i} merges a final report that contains all healthy devices
in the network. This final report is encrypted under $\devicekey_i$ and transmitted
to $\mathcal{O}$, who verifies the report, as described in the next paragraph.

We note that the attestation must be completed in time $\tattack$
or $\mathcal{O}$ has to periodically check the presence of \device{i}
during attestation.
Otherwise, \Adv can physically tamper with \device{i} to extract an aggregate
and induce attests of physically compromised devices.
Figure~\ref{fig:protocol:attestation}
illustrates the attestation phase in a network with $6$ healthy devices and one
adversary device \device{A} that was physically compromised.

\para{Report Aggregation and Merging}
An aggregated attestation report consists of two parts. The first part contains
a description of all device identifiers that are in the aggregate.~The second part
consists of the aggregated attests.
For a small number of devices, the description is a list of device identifiers,
else it is an $n$-bit vector, where a one at position
$k$ indicates that \device{k} is contained in the aggregate.
The attests themselves are aggregated by XOR-ing all
individual attests. 
Multiple attestation reports are aggregated by merging their device descriptions and XOR-ing their aggregated attests.


When attesting the overall network state, the attestation report consists of only
the aggregate, as a device identification is not required.
This decreases the size of the report significantly (\S~\ref{evaluation:static}).
Therefore, to increase efficiency, it is useful to run the attestation
with precise device identification only, if an attestation of the overall network state fails.





\para{Report Verification}
Given a device description, $\mathcal{O}$ recomputes the attests for all devices,
whose id is contained in the description. Given no description, $\mathcal{O}$ recomputes
the attests for all devices.
If the recomputed aggregate equals the reported aggregate and if at least
$n/2$ attests are included in the report, then the report is assumed to
be valid. Only then, all attested devices are assumed to be healthy and
the verification returns a bit vector, where a zero/one at position $k$
indicates that \device{k} is compromised/healthy.




\subsection{Security Analysis}
\label{sec:protocol:security}
Intuitively, an attestation protocol is secure, when the
network operator $\mathcal{O}$ will testify a healthy system state, if not
a single device has physically been compromised. We refer to such
an attestation scheme as \emph{non-informative} secure.
Moreover, an \emph{informative} secure attestation protocol allows $\mathcal{O}$ to
distinguish between healthy and compromised devices.
We follow the idea of Asokan et al.~\cite{asokan2015seda} and
prove the security of our protocol by an adversarial experiment
$\ExpA(k)$. In this experiment, the adversary $\Adv$ is given access
to a network of $n$ initialized devices $\Net$ that execute
the heartbeat and attestation protocol.
$\Adv$ can interact with all devices
according to the attacker model presented in \S~\ref{pre:adv}.
%
%
%
Moreover, we assume any adversary \Adv to be computationally bound (PPT).
Hence, \Adv is able to interact a polynomial number of times $k$
with devices in the network (and the authenticated encryption scheme).
Furthermore, $\Adv$ is allowed to trigger
and observe attestations by $\mathcal{O}$.
After at most $k$ interactions, a final attestation is initiated by $\mathcal{O}$.
The output of $\ExpA(k)$ is then a bit vector returned by $\mathcal{O}$ after
verification of the final request. A bit vector with only zeros
indicates a compromised network, whereas every bit set to one indicates
a healthy device, cf. \S~\ref{sec:protocol:attestation}.
We capture the intuitive idea of secure attestation in
the following definition.
\begin{definition}
\textbf{Secure Attestation Scheme.}  An network attestation scheme
for $n$ devices is secure if
\[\Pr[\ExpA(k) = 1^n] \leq negl(k)\]
for any PPT \Adv and $0 < c < n$, where c is the number of compromised devices.
An attestation scheme is informative and secure if
\[ \Pr[\ExpA(k)[j] = 1] \leq negl(k) \] for any PPT \Adv and every compromised
device $\device{j}$, where [j] is the j'th bit in the result vector
and the total number of compromised devices $c$ is 
less than $n/2$.
\end{definition}
Note that the definition of a non-informative secure attestation scheme
is similar to the definition given in \cite{asokan2015seda}, which is
defined without device identification in mind.

\para{Security of \ourname}
The security of \ourname is summarized
in Theorem~\ref{theorem:security}.

\begin{theorem}
\label{theorem:security}
\ourname is an informative and secure attestation protocol when the
length of a heartbeat period $\Tslot$ is at most $\tattack/2$, assuming 
security of the PRNG and authenticated encryption
scheme that guarantees confidentiality (IND-CPA) and
authenticity (INT\--CTXT).
\end{theorem}

In the following paragraphs,
we sketch a proof to show that \ourname is an informative secure
attestation scheme.
The sketch is split in two parts. First, we sketch a proof for 
Theorem~\ref{theorem:heartbeat}, which formalizes the security
of the heartbeat protocol, before arguing the security of
the full protocol.


\begin{theorem}
\label{theorem:heartbeat}
Any PPT \Adv is unable to gain access to any heartbeat $\hb_t$, which is used to
secure the communication in time period $t$, before
time period $t+1$, assuming $\Tslot < \tattack/2$, security of the PRNG, secure channels between devices and an authenticated encryption scheme that guarantees IND-CPA and INT-CTXT.
\end{theorem}

Intuitively, the security of the heartbeat protocol is achieved
by using an interactive protocol that requires the receiving device to
prove its knowledge about the current heartbeat to the sending device.
Only then, the next heartbeat is exchanged.
This active participation makes it impossible
for offline devices to follow the continuous `stream' of heartbeats.

\para{Proof Sketch - Heartbeat}
We observe that no two heartbeats are linked. 
Hence, it is impossible to derive any $\hb_{t}$
from $\hb_{1}$,$\hb_{2},\dots,\hb_{t-1}$ without breaking the security
of the PRNG. Moreover, assuming synchronized clocks, every healthy device stores
at most two heartbeats in any time period $t$, namely $\hb_{t-1}$,
$\hb_{t}$ or $\hb_{t},\hb_{t+1}$. When compromising a single device in time
period $t$ and assuming an attack time of $\tattack \geq 2 \cdot \Tslot$,
the attack will be successful not earlier than in time period $t+2$.
The \ac{TEE} of the compromised device will then leak 
at most heartbeat $\hb_{t+1}$, but no later heartbeats, as 
these are not present in the \ac{TEE}.
We observe that with any attack time $\tattack < 2 \cdot \Tslot$, $\Adv$ would be able
to compromise a device without missing a single heartbeat period, and thus
render the protocol insecure.

We show that $\Adv$ is unable to gain access to the current heartbeat
by interacting with healthy devices without breaking the security
of the authenticated encryption scheme. During the heartbeat exchange,
all messages sent between two devices \device{i} and \device{j} are encrypted with a \emph{session}
key that is the XOR of the pairwise channel key $\key_{ij}$ and the current heartbeat $\hb_t$ at time $t$.
Thus, the session key is only known to \device{i} and \device{j} at time $t$.
We observe that with access to only one (or none)
of the two keys, $\Adv$ is unable to create or to decrypt a message that
is accepted by \device{i} or \device{j} without breaking the INT-CTXT and 
IND-CPA security of the encryption scheme. Hence, even when compromising
further devices and extracting (past) heartbeats, \Adv is unable to
decrypt any past or future communication between \device{i} and \device{j}, as \Adv
is missing the pairwise channel key $\key_{ij}$. Similarly, after compromising
a device and gaining access to all channel keys, \Adv is still
missing  the current heartbeat to construct the session key, required to
interact with neighboring devices.
%
The same arguments hold for all messages sent between devices in the
aggregation protocol, since they are all encrypted using the pairwise session key.



\para{Proof Sketch - Attestation} 
The attest of a single device $\device{i}$ is
the encryption of the timestamp $ts$, issued by $\mathcal{O}$, under $\device{i}$'s
device key $\devicekey_i$. Thus, \Adv is only able to forge an attest for
a healthy $\device{i}$ with non-negligible probability when being able to
break the IND-CPA security of the encryption scheme.
Yet, to win $\ExpA$, \Adv has to report at least $n/2$ (informative) or
$n$ (non-informative) valid attests, while being allowed to only compromise up to $c < n/2$ 
or $c<n$ devices. Consequently, since \Adv is unable to forge an attest for a healthy
device with non-negligible probability, \Adv has to merge the attests of compromised
devices with attests created by healthy devices.

During the actual attestation protocol, two cases can be distinguished.
First, the device
$\device{i}$ that $\mathcal{O}$ approaches for the attestation
is compromised. In this case, $\Adv$ can create an attestation report
for all compromised device. However, without access to a valid heartbeat 
and thus session key, $\Adv$
can only create a valid attestation request message $\msg_{att}$
with non-negligible probability, when breaking the INT-CTXT security
of the encryption scheme. Hence, no healthy device will contribute an
attest.
Similar, in the second case, where $\mathcal{O}$ first approaches a
healthy device, \Adv is, for the same argument as described above,
unable to decipher or induce any message in the
attestation protocol between healthy device.
Furthermore, the security of a XOR aggregation scheme, as used here,
is shown in~\cite{katz2008aggregate} and consequently,
\ourname is non-informative secure, when only accepting
a complete aggregation report that includes the attests of all devices. 
Furthermore, it is informative secure, when accepting 
reports with at least $n/2$ attests,
because attests can be attributed towards their device id.
Finally, we remark that the `honest majority' assumption $c<n/2$ is required, as
otherwise a dishonest majority could fake a healthy systems state.




\section{Protocol Extensions}

\label{extensions}
In the following, we present three significant extensions to \ourname. First, we make the heartbeat transmission phase more robust against failures (\S~\ref{extensions:leaderelection}). Next, we extend \ourname to verify the integrity of the software on all devices in the network (\S~\ref{extensions:software}). Finally, we propose an extension that allows efficient attestation in highly dynamic and disruptive network topologies (\S~\ref{extensions:dynamic}).



\subsection{Leader Election Protocol}
\label{extensions:leaderelection}
The leader election phase extends the heartbeat transmission phase, to make it more robust against failures. In particular, devices that fail to receive the current heartbeat elect a new leader device that takes over the tasks of the previous leader, i.e., the periodic emission of a new heartbeat. In this way, the heartbeat protocol is able to recover from device outages, network partitioning, or targeted denial of service attacks.

The leader election protocol is initiated by every device that fails to receive the heartbeat within a time $\Tslot_{hb}$ that is shorter than the heartbeat period $\Tslot$ $(\Tslot_{hb} < \Tslot)$. Devices execute the leader election protocol inside their \ac{TEE} and use the remaining leader election time $\Tslot_{le} = \Tslot - \Tslot_{hb}$ to determine the device with the smallest id, which then becomes the new leader device
(bully algorithm). For this purpose, devices initially generate their own heartbeat and then announce this heartbeat together with their device id to all neighboring devices. Devices store the smallest device id that they received in the leader election phase, including the corresponding heartbeat. Whenever a device updates its smallest received id and heartbeat, it broadcasts both to their neighboring devices. Thus, the new smallest id and heartbeat are quickly propagated in the network. A device recognizes itself as the new leader device, if it only receives messages from devices with higher device ids.~Note that the original leader has the smallest id in the entire network, hence, the protocol also tolerates a return of the original leader. In Appendix~\ref{appendix:leaderelection}, we formalize the leader election protocol, describe it in more detail, and demonstrate its security.

\subsection{Attestation of Software Integrity}
\label{extensions:software}
In order to attest the correct and safe operation of all devices in the network, it is crucial to ensure that devices are in a trustworthy software state, free from malicious or broken software. For this purpose, we propose that the network operator $\mathcal{O}$ defines a set of trustworthy software states $tss$ in the attestation request, when initiating an attestation of the network. $Tss$ specifies all software configurations that are permitted by $\mathcal{O}$, e.g., because they represent the correct and most recent software states. When devices perform the attestation protocol, they invoke the execution of a software integrity measurement function in their \ac{TEE}. This function measures the integrity of installed software and compares these measurements to the reference values specified in $tss$. In this way, each device determines whether it is in a trustworthy or untrustworthy software state. Devices being in an untrustworthy software state immediately abort the attestation phase and instead execute a recovery routine that allows the device to restore a trustworthy software state, e.g., via secure code updates~\cite{kohnhauser2016secure}. Since untrustworthy devices do not participate in the execution of the attestation protocol, $\mathcal{O}$ receives a report which exclusively contains devices that are in a trustworthy software and uncompromised hardware state. In Appendix~\ref{appendix:softwareintegrity}, we extensively explain changes that need to be done to the enrollment phase and the attestation protocol to enable such a hybrid attestation. Furthermore, we discuss the security of the extension.

\subsection{Attestation of Dynamic Networks}
\label{extensions:dynamic}

\para{Approach}
The attestation protocol in \ourname (\S~\ref{sec:protocol:attestation}) arranges a spanning tree, which allows for an efficient aggregation and transmission of the attestation report to the network operator $\mathcal{O}$. This approach works efficiently as long as the network topology stays static during attestation, for instance, as devices in the network only move as a whole (herd mobility) or within local limits (micro-mobility). However, in dynamic network topologies with highly mobile devices and frequent link disruptions, it is impractical to maintain a spanning tree topology. In such networks, communication with a parent device could introduce a significant delay
or become highly inefficient, as the parent device could move away. Even worse, the parent device may be temporarily out of range and thus be disconnected from the network. 

Therefore, instead of routing the attestation along a virtual topology, we propose a distributed (greedy) aggregation, where attestation reports are collected and aggregated by all devices in the network. Thus, after $\mathcal{O}$ initiates the attestation protocol, each device first generates its own attestation report, stores this report, and broadcasts it to all neighboring devices. When a device receives an attestation report, it merges this report with its stored report. On observing new attests, the device broadcasts the updated report to all its neighboring devices. In this way, all devices in the network eventually store the same attestation report and $\mathcal{O}$ can obtain the attestation result from an arbitrary device in the network.

To reduce the communication complexity, an aggregation scheme for the above mentioned approach must allow to merge multiple reports with intersecting attests into one. This requirement renders the aggregation function described in \S~\ref{sec:protocol:attestation} inapplicable, because its XOR operation risks the removal of intersections of attests from the aggregate. 
Because of this and following the analysis of aggregation protocols in \S~\ref{relatedwork}, we present a novel aggregation scheme for dynamic networks that is particularly tailored to the application scenario.





\para{Secure \& Efficient Attestation Report Aggregation}
The here proposed scheme achieves statistical security and is slightly less powerful
than the spanning tree aggregation scheme, as it allows an adversary to
compromise at most $c < n/2 - s$ devices, with $2^{-s}$ being the statistical
security level.
In our scheme, an attestation report also consists of two parts, namely
the device description and the secure aggregate itself. 
The device description is a $n$-bit vector where a bit is set for every
device included in the aggregate. The aggregate consists of an
$n_s =(n+s)$-bit vector, where a single bit indicates the attest of a
device. A device \device{i} that receives an attestation request with
timestamp $ts$, creates its own attest using a collision
resistant cryptographic hash function $\hash$ by computing
$a = \hash(dk_i || ts)$ in its TEE. Subsequently, \device{i} 
sets a bit at position $i$ in the device identifier as well as
a bit at offset $compress(a)$ in the secure aggregate, where
$compress$ is a function that reduces the hash value to a value of
length $n_s$ bits. Note that $compress$ does not need to be 
cryptographically secure, but it should achieve a close to uniform 
output distribution for uniformly distributed input.
All other bits in both vectors are set to~$0$.
In order to merge multiple attestation reports, a device
computes the bit-wise OR of all attestation reports. This can be done
very efficiently and allows to aggregate reports with intersections of
devices. 
Both the secure aggregate and the list of device identifier could be
compressed, for instance, by using a run-length encoding. Nevertheless,
even without compression, a very short attestation report
is achieved with a length of only $2n+s$ bits, e.g., 266 bytes for
1000 devices and a security level of $s=128$ bit, which is a significant
improvement over a na\"{i}ve concatenation of attests that requires more
than $16\text{k}$ bytes. Even though, \Adv has a good chance to guess
a small number of attests correctly, the security of the scheme is based
on the hardness to guess (at least) $s$ attests correctly.
A detailed security analysis of this scheme is given in
Appendix~\ref{appendix:aggregation}.

\vspace{-2mm}
\section{Evaluation}
\label{evaluation}

Next, we evaluate \ourname (\S~\ref{protocol}) and its three protocol extensions (\S~\ref{extensions}). In \S~\ref{evaluation:implementation}, we describe our setup, give details of the implementation, and present our measurements. Then, we report on our network simulation results for both static (\S~\ref{evaluation:static}) and dynamic network topologies (\S~\ref{evaluation:dynamic}).

\subsection{Implementation \& Measurements}
\label{evaluation:implementation}

\para{Setup}
We implemented our protocol on Stellaris EK-LM4F120XL microcontrollers. The Stellaris is a low-cost embedded system from Texas Instrument which features an 80 MHz ARM Cortex-M4F microprocessor and 256~kB of Flash memory. To enable wireless mesh connectivity based on the ZigBee standard, we equipped the Stellaris microcontrollers with Anaren's CC2530 BoosterPacks.


\para{Cryptographic Runtime Measurements}
We im\-ple\-men\-ted the hash function using SHA-512 and employed AES in Galois/Counter Mode (AES-GCM) as an authenticated encryption scheme. For the key exchange, we used Elliptic Curve Diffie-Hellman (ECDH) with Curve\-25519~\cite{bernstein2006curve25519}. Table~2 shows an excerpt of our cryptographic runtime measurements on the Stellaris microcontroller.

\begin{table}[t]
\centering
\renewcommand{\arraystretch}{1.2}
\begin{tabular}{l l l }
\toprule
\textbf{Algorithm} & \textbf{Function} & \textbf{Runtime} \\
\midrule
ed25519 & \textsf{genKey()} & 18 ms \\
& \textsf{keyExchange()} & 48 ms \\
\midrule
AES-128-GCM & \textsf{encrypt($16$ $bytes$)} & 0.1 ms \\
& \textsf{encrypt($1024$ $bytes$)} & 1.8 ms \\
& \textsf{decrypt($16$ $bytes$)} & 0.1 ms \\
& \textsf{decrypt($1024$ $bytes$)} & 1.8 ms \\
\midrule
SHA-512 & \textsf{hash($16$ $bytes$)} & 0.4 ms \\
& \textsf{hash($1024$ $bytes$)} & 3.1 ms \\
& \textsf{hash($30720$ $bytes$)} & 81.9 ms \\
\bottomrule
\end{tabular}
\label{table:cryptoperformance}
\caption{Crypto Runtime Performance on the Stellaris.}
\end{table}

We would like to stress that our implementation is based on platform independent and unoptimized C code.\footnote{\small We used SUPERCOP's ed25519 implementation (\url{https://ed25519.cr.yp.to/software.html}) and SharedAES-GCM (\url{https://github.com/mko-x/SharedAES-GCM}).}~Assembler optimized code for low-end embedded systems can improve the performance of cryptographic operations by orders of magnitudes~\cite{dull2015high, wenger20138}.~We presume that similar performance improvements are also possible on the Stellaris.


\para{Network Runtime Measurements}
For unicast messages between two neighboring devices in the mesh network, we measured an average throughput of 35.0 kbps on the application layer. Although the theoretical maximum throughput in ZigBee networks is 250 kbps, other performance evaluations revealed similar performance losses in reality~\cite{burchfield2007maximizing}. In addition, we measured an average end-to-end delay between two neighboring devices of 13.5 ms with the smallest message size and 18.5 ms with the biggest message size.


\para{Memory Costs}
In our implementation, devices store their own ECDH key pair (64 bytes), the current and the next heartbeat (each 16 bytes), the leader device id (4 bytes), $k$ secure channel keys and device ids (each 20 bytes), and a timestamp (4 bytes). The number $k$ of stored secure channel keys can be adjusted to the particular memory requirements, since devices can establish channel keys right away by performing an ECDH key exchange with a neighboring device (\S~\ref{sec:protocol:heartbeat}). Additionally, devices need to temporarily store data: the public key of a neighboring device (32 bytes) during key exchange, a session key (16 bytes) during heartbeat transmission, and the attestation report during attestation. The size of the attestation report is dependent on the number $n$ of devices and the usage of the dynamic network extension. If it is used with a security level of 128 bit, the report amounts to $n/4 + 16$ bytes, if not, to $n/8 + 16$ bytes. However, as reports can be compressed using run-length encoding, their actual size is much lower for most devices in the network. In total, devices require $104 + k \cdot 20$ bytes of permanent storage and at most $n/4 + 16$ bytes of temporary storage.


\subsection{Simulation Results for Static Networks}
\label{evaluation:static}
\smallskip\noindent\textbf{Setup.}
We first evaluated our protocol in static network topologies, where all devices are connected and stationary. Thus, there are no link breaks or abrupt delays in the network communication. We used \texttt{ns 3.25}~\cite{ns3} to simulate a homogeneous network with ten to multiple million Stellaris devices. Following the typical evaluation methodology in scalable attestation protocols~\cite{ambrosin2016sana, asokan2015seda, ibrahimdarpa}, we implemented our protocol on the application layer and used computational and network delays based on our measurements (see \S~\ref{evaluation:implementation}).





\begin{figure}[t]
	\centering
	\includegraphics[width=0.48\textwidth]{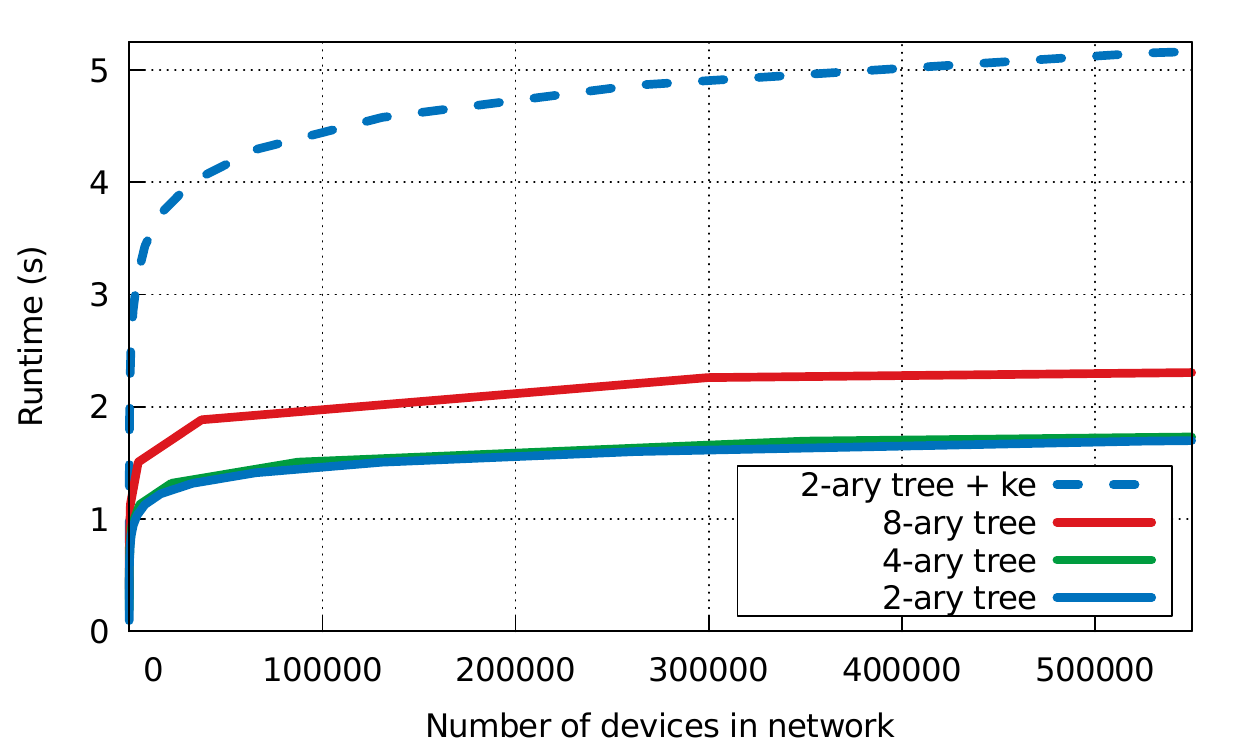}
	\caption{Heartbeat protocol runtimes protocol runtimes in various static topologies. The dotted line shows the runtime of the first heartbeat with the initially needed key exchanges.}
	\label{fig:heartbeat:static}
\end{figure}

\para{Heartbeat Protocol Runtime}
We simulated the runtime of the heartbeat protocol in various topologies. Figure~\ref{fig:heartbeat:static} shows the runtime for a binary, 4-ary, and 8-ary tree topology with up to 550.000 devices, where the heartbeat leader device is located at the root of the tree. The figure demonstrates that in tree network topologies, protocol runtime increases logarithmically with the number of devices in the network. Under these conditions, the heartbeat protocol achieves an outstanding performance, requiring less than 2.3 seconds to reach 500.000 devices in an 8-ary-tree and less than 1.7 seconds in a binary tree topology. Even with multiple million devices, runtime remains below 2 seconds in a binary-ary tree topology. Only the first run of the heartbeat protocol in the network requires little more time, since neighboring devices initially need to exchange public keys and perform key exchanges to establish shared secrets. Yet, even with the additional key exchanges runtime remains below 5.1 seconds for more than 500.000 devices.



\para{Attestation Protocol Runtime}
We configured the attestation protocol to use the software attestation extension (\S~\ref{extensions:software}) and thus to attest the hardware and software state of all devices in the network. To verify the integrity of installed software, devices compute a SHA512 digest over a 30 kB software and compare the digest to an expected value that is specified in the attestation request. 
For attestation we used the spanning tree attestation approach (\S~\ref{sec:protocol:attestation}).


\begin{figure}[t]
	\centering
	\includegraphics[width=0.48\textwidth]{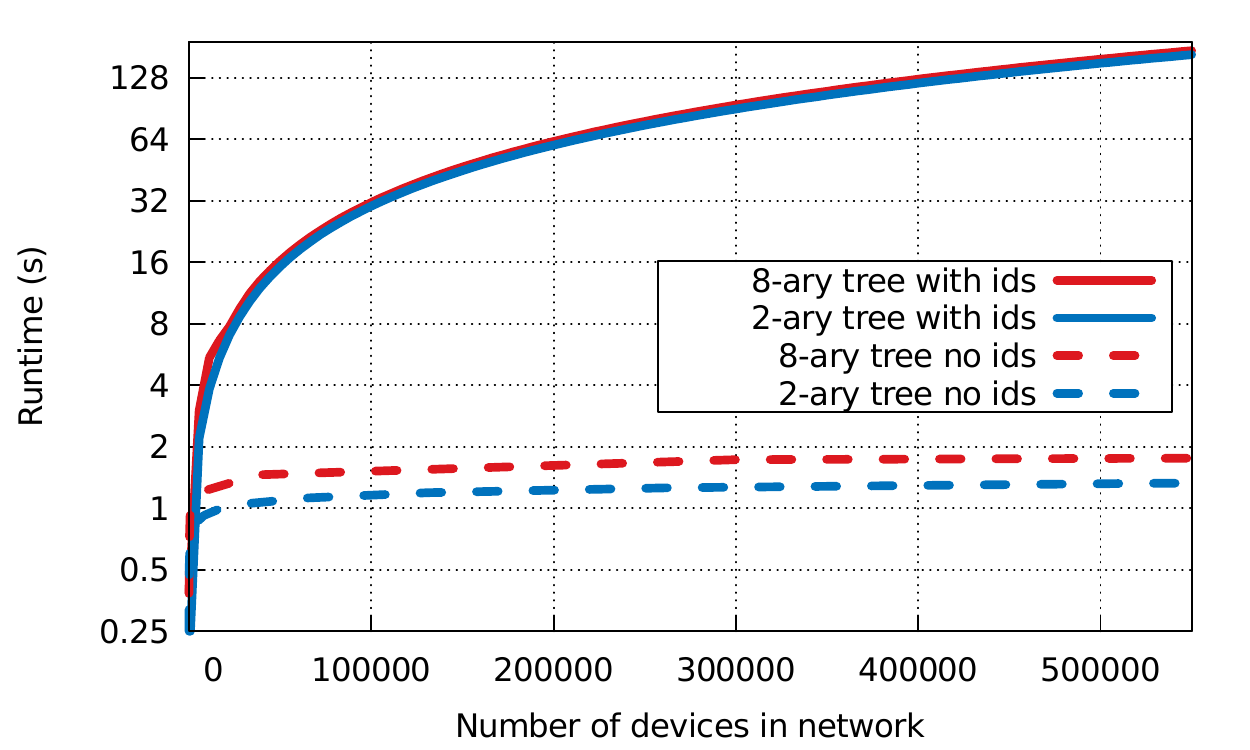}
	\caption{Attestation protocol runtimes in two static topologies with and without sending device identifiers in the aggregate.}
	\label{fig:attestation:static}
\end{figure}

Figure~\ref{fig:attestation:static} shows the runtime for a binary and 8-ary tree topology with up to 550.000 devices, where the network operator is located at the root of the tree. Additionally, we varied the type of the attestation report, containing either the precise ids of healthy devices (solid lines) or the state of the overall network (dashed lines). The figure demonstrates that reporting precise device identifier introduces a notable overhead. When reporting the overall network state, attestation runtime increases barely with the number of devices in the network, remaining below 2 seconds even for networks with multiple million devices in almost any tree topology. Yet, when reporting precise device ids, runtime increases to more than 152 seconds for 500.000 devices due to the large size of the attestation report, which increases proportionally with the network size. Nevertheless, we consider that 2.5 minutes is an acceptable timeframe to obtain a report that precisely lists which devices 
are in a compromised state.


\para{Communication Costs} 
During heartbeat transmission, all devices, except for the leader device, receive  $msg_{new}$ (1~byte), send $msg_{req}$ (17 bytes), and receive $msg_{hb}$ (17 bytes) to obtain the newest heartbeat, using a one byte message identifier. 
If devices need to (re-)establish a secure channel key, they need to mutually exchange their public keys, which causes an additional message overhead of 32 bytes. For instance, in a binary tree topology, devices transmit in total $104$ bytes, or $296$ bytes with the initial key exchange, in each heartbeat transmission period. 

During the execution of the attestation protocol, all devices receive one $msg_{V}$ (17 bytes) or $msg_{att}$ (17 bytes).
Also, devices send a $msg_{att}$ to all neighbor devices that have not yet received $msg_{att}$ and afterwards receive a $msg_{agg}$ from them ($\le n/8 + 16$ bytes). If the device's software integrity is attested (\S~\ref{extensions:software}), $msg_{V}$ and $msg_{att}$ contain the set of trustworthy software states $tss$, in our evaluation a $64$ bytes hash digest. In short, assuming a binary tree topology and $n = 1000$ devices, during a run of the attestation protocol, each non-leaf device transmits at most $666$ bytes and each leaf device $222$ bytes.




\para{Summary}
We demonstrated that our protocol is highly efficient in static network topologies. In comparison to the previously best attestation protocol that is secure against physical attacks~\cite{ibrahimdarpa}, we reduce the number of transmitted messages per time period from $\mathcal{O}(n^2)$ to $\mathcal{O}(n)$. 
To illustrate this advantage, in binary-tree topologies our approach is $27$ times faster with $2000$ devices and $3800$ times faster with $500,000$ devices when interpolating their results. The comparison already considers the fastest variant presented in~\cite{ibrahimdarpa}, which requires each device to store and manage $n$ symmetric keys. In our protocol, devices must only store the keys of neighboring devices, e.g., $3$ in a binary tree topology.


When attesting the state of the entire network, both protocols (\cite{ibrahimdarpa} and \ourname) show a runtime that scales logarithmically with $n$. Nevertheless, in contrast to \cite{ibrahimdarpa}, \ourname also allows to determine the ids of compromised devices with low overhead even in larger networks.


\subsection{Simulation Results for Dynamic Networks}
\label{evaluation:dynamic}
\para{Setup}
We further evaluated our protocol in highly dynamic and disruptive networks to investigate its robustness in complex scenarios. To model device mobility, we randomly deployed devices in a 1000m x 1000m square area and applied a random waypoint mobility model, which is commonly used in literature on absence detection~\cite{conti2009mobility, di2010securing}. Consequently, each device repeatedly selects a random speed as well as a random destination within the area and then moves towards the destination at the selected speed. The random device movement causes the network to be constantly partitioned, especially for sparse networks. In order to investigate effects like link disruptions, varying network delays, and signal interference that emerge due to the movement of devices, we modeled an 802.15.4 physical and medium access control layer using the \texttt{ns-3.25 lr-wpan} module. Modeling both layers as well as device mobility requires a lot of computational power. This is a known issue in MANET simulations, which leads to huge simulation runtimes~\cite{bilel2012hybrid}. For these reasons, we were only able to run simulations with a few hundred devices. Nevertheless, as we will show in this section, the main hurdle of our protocol is to perform well in sparse networks. Scalability of our approach in dense networks, where all devices are permanently interconnected, is shown in the previous section. In addition to the above mentioned simulation parameters, we set the wireless communication range to 50m ($50\%$ of the distance specified in the ZigBee standard), the device speed to a random value between 5 and 15 m/s, and the heartbeat as well as the leader election period to 2.5 minutes (detecting physical attacks that require more than 10 minutes).



\begin{figure}[t]
	\centering
	\includegraphics[width=0.48\textwidth]{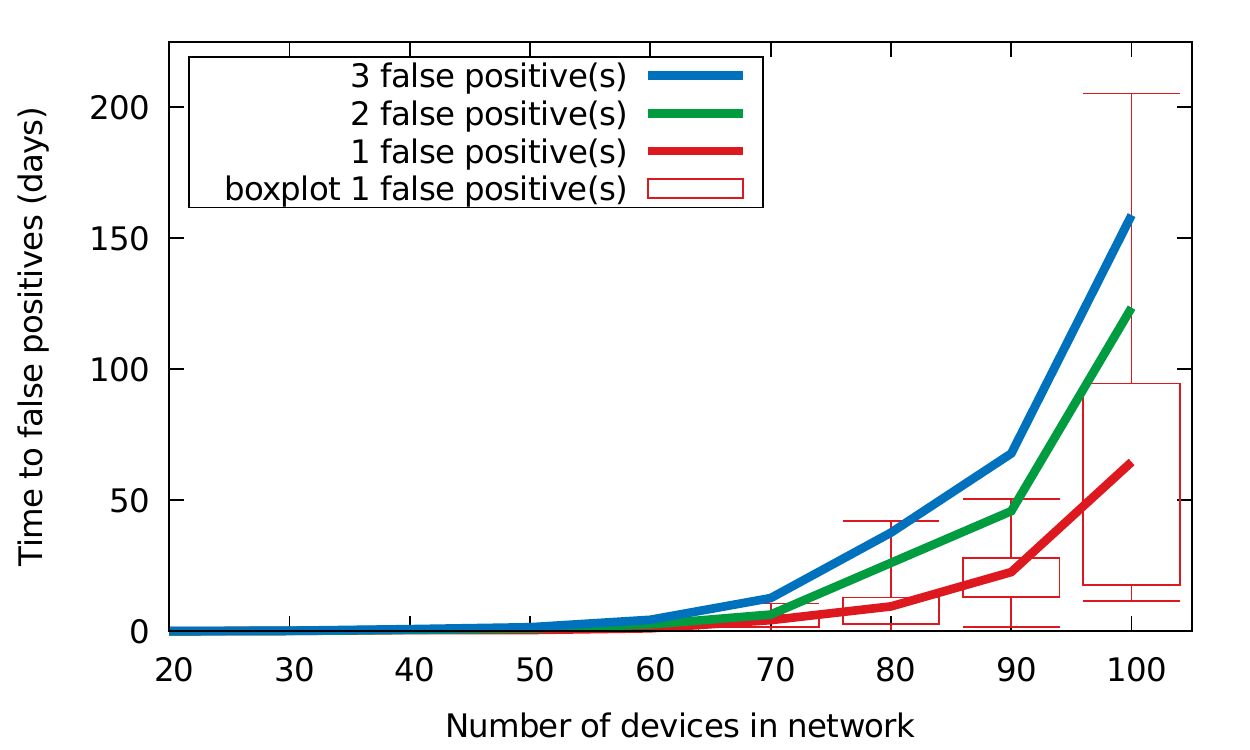}
	\caption{Heartbeat protocol average runtime in a dynamic topology until false positives occur. The boxplot shows the distribution of simulation results for the single false positive case.}
	\label{fig:heartbeat:dynamic:boxwhisker}
\end{figure}

\para{Heartbeat Protocol Robustness}
We investigated the robustness of the heartbeat protocol in worst cases, which are highly dynamic and disruptive network topologies. In particular, we examined the time until the protocol produces false positives, i.e., healthy devices that are regarded as physically compromised, because they did not receive the heartbeat on time. Figure~\ref{fig:heartbeat:dynamic:boxwhisker} illustrates the average runtime of the heartbeat protocol until a certain amount of false positives occur. The figure shows that the number of devices in the network has a vital influence on the robustness of the heartbeat protocol. Since devices move completely at random, the network must be sufficiently dense so that devices meet each other frequently enough to exchange the newest heartbeat on time. In fact, there is an exponential correlation between robustness and device density, which causes the average error-free heartbeat runtime to quickly increase from 2.4 days for 60 devices to weeks with more than 90 devices. To illustrate the sparseness in this scenario, 60 statically connected devices could cover at the maximum $29\%$ of the area and 90 devices $43.4\%$. Nevertheless, as shown by the boxplot, the runtime between multiple simulation results differ widely. This makes it hard to guarantee robustness for sparse network scenarios. Investigating the false negatives, we identified the main cause in the random movement of devices. Commonly, a single device hides away, i.e., does not encounter other devices, and thus has no chance to receive the newest heartbeat on time. This cannot be prevented by faster computations or smaller communication delays in our protocol, but only be increasing the duration of the heartbeat phase.
We also observed that this hiding of a single device has barely any cascading effect on other devices. Hence, as shown in Figure~\ref{fig:heartbeat:dynamic:boxwhisker}, if tolerating a minimal amount of false positives, significant longer protocol runtimes are possible.





\begin{figure}[t]
	\centering
	\includegraphics[width=0.48\textwidth]{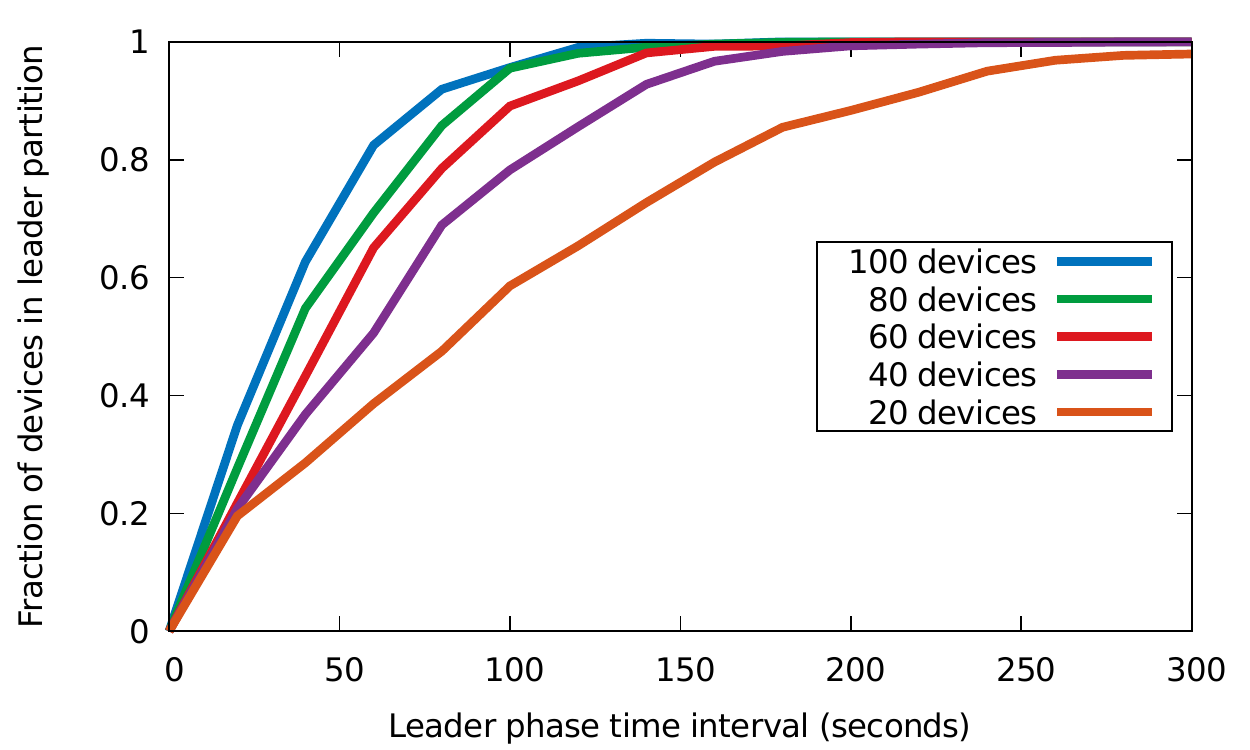}
	\caption{Efficiency of leader election in heartbeat protocol.}
	\label{fig:heartbeat:dynamic:leader}
\end{figure}

Next, we analyzed the effectiveness of the leader election extension by simulating an outage of the heartbeat leader device. Figure~\ref{fig:heartbeat:dynamic:leader} shows the largest fraction of devices that agreed on a common new leader with an increasing time interval for the leader election phase. It again illustrates the importance of the network density. In dense networks, leader election information can spread faster and thus reach more devices in shorter time. Nevertheless, even in relatively sparse networks with 60 devices, a time interval of 150 seconds is on average sufficient to let all functioning devices agree upon a new common leader. This also highlights our robustness against targeted \ac{DoS} attacks, where an adversary attempts to disrupt the heartbeat protocol by breaking the heartbeat leader device.

\begin{figure}[t]
	\centering
	\includegraphics[width=0.48\textwidth]{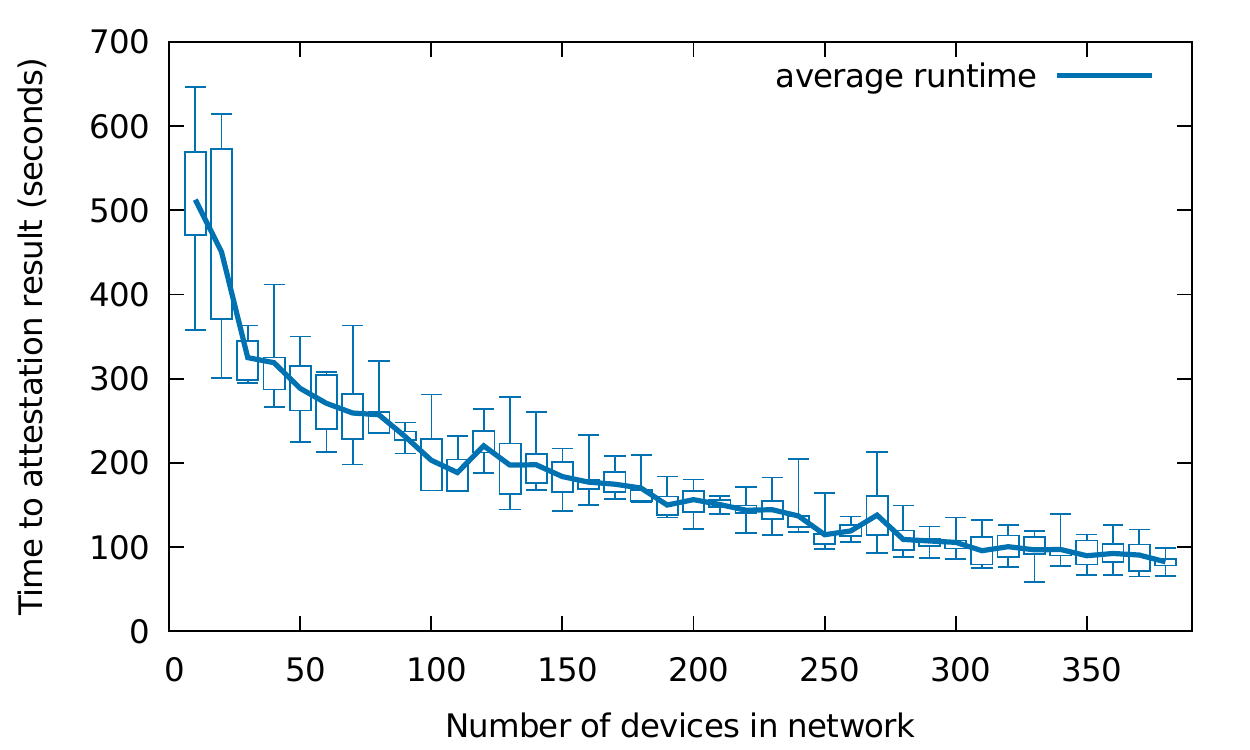}
	\caption{Runtime of attestation protocol in dynamic topologies with an increasing number of devices.}
	\label{fig:attestation:dynamic:incnodes}
\end{figure}

\para{Attestation Protocol Runtime}
For the evaluation of the attestation protocol in dynamic networks, we used our dynamic attestation extension (\S~\ref{extensions:dynamic}) with a statistical security level $s$ of $128$ bits. Figure~\ref{fig:attestation:dynamic:incnodes} shows a boxplot of the elapsed time between the emission of the attestation request and the moment when all devices in the network store the final attestation result, i.e., a report that contains all devices, for an increasing number of devices in the network. With an increasing network density, attestation reports spread faster and the overall attestation protocol runtime decreases. However, this effect is not as distinct as with the heartbeat protocol, where we observed an exponential correlation between protocol performance and network density. This is because, in contrast to the heartbeat protocol, where a single message is flooded in the network, each device must contribute with a message, i.e., its individual attestation report, to a global attestation aggregate.~Furthermore, the size of the attestation report increases proportional with the network size, though, being reasonably small for common network sizes (e.g., 2.5kB for $10,000$ devices). Figure~\ref{fig:attestation:dynamic:incnodes} also shows that the runtime of the attestation protocol varies little. This guarantees that the final result is with high probability reached within a certain time frame, e.g., 5 minutes for 100 devices.


\para{Communication Costs}
Message costs in dynamic network topologies are, except for the attestation report, the same as in static network topologies (\S~\ref{evaluation:static}). However, due to link failures, some messages are transmitted more often in dynamic topologies. In our simulations, we varied the network size between $40$ and $100$ devices and let devices actively poll their neighbors for the newest heartbeat after 10 seconds. Our results revealed that, depending on the network size, each device sends on average $19.4$ to $21.0$ $msg_{new}$ (poll heartbeat) and $1.04$ to $1.12$ $msg_{req}$ as well as $msg_{hb}$ messages. Hence, in total, devices transmit on average $114$ bytes in each heartbeat transmission phase. Compared to static network topologies (\S~\ref{evaluation:static}), this is less than $10\%$ communication cost overhead.

Nevertheless, because the attestation result is distributed to all devices, the actual attestation consumes considerably more communication in disruptive networks. Conducting the same simulations as described above, we observed that each device exchanges on average $12.4$ attestation reports in networks with 40 devices, $16.0$ with 60 devices, $18.9$ with 80 devices, and $21.5$ with 100 devices. Each exchange requires a device to send one $msg_{att}$ and receive one $msg_{agg}$. Note that the dynamic attestation report in $msg_{agg}$ has a size of at most $n/4 + 16$ bytes. Thus, in total, devices transmit on average $1375$ bytes in networks with 100 devices, which is 4.2 times more than in static network topologies.




\para{Summary}
We showed that our heartbeat and attestation protocols are robust and efficient, even in highly partitioned and unpredictably changing network topologies. In fact in an exemplary low connectivity scenario, with a maximum possible area coverage of $43\%$ for randomly moving devices, the heartbeat protocol still runs on average 65 days without producing a single false positive with $\tattack=10$ minutes. We further illustrated the effectiveness of the leader election protocol, by completely recovering networks from device outages in less than 130 seconds in the same setting. Finally, we demonstrated the robustness of our attestation protocol in dynamic networks and showed that its performance is dominated by network connectivity as opposed to the protocol's communication complexity.





\section{Conclusion \& Future Work}
\label{conclusion}
We presented the first scalable attestation protocol \ourname for mesh networked embedded devices that is resilient to physical attacks. Compared to existing solutions, our protocol reduces the number of transmitted messages per time period from $\mathcal{O}(n^2)$ to $\mathcal{O}(n)$, thus scaling to millions of devices and outperforming existing solutions by orders of magnitude. In addition to attesting the overall state of the network, \ourname is able to precisely identify devices that run compromised software or have been physically manipulated. 
We demonstrated that our protocol is robust and efficient, even in very dynamic topologies, as it can perform an attestation or recover from device outages within minutes.

In future work we plan to investigate our protocol in specific network application scenarios, such as drone-based delivery systems or wireless sensor networks. Moreover, we want to make use of MANET simulators that are optimized for scalability and/or parallelism, in order to be able to simulate thousands of moving devices.






\bibliographystyle{abbrv}
\bibliography{sigproc}

\begin{thebibliography}{10}

\bibitem{ns3}
{NS-3 Network Simulator}.
\newblock \url{https://www.nsnam.org/}.

\bibitem{gartner2015iot}
{Gartner Says 6.4 Billion Connected "Things" Will Be in Use in 2016, Up 30
  Percent From 2015}, 2015.
\newblock \url{http://www.gartner.com/newsroom/id/3165317}.

\bibitem{abera2016c}
T.~Abera~et al.
\newblock {C-FLAT: Control-FLow ATtestation for Embedded Systems Software}.
\newblock In {\em ACM CCS'16}.

\bibitem{agrwal15:programIntegrity}
S.~Agrawal~et al.
\newblock {Program Integrity Verification for Detecting Node Capture Attack in
  Wireless Sensor Network}.
\newblock In {\em ICISS'15}.

\bibitem{ambrosin2016sana}
M.~Ambrosin~et al.
\newblock {SANA: Secure and Scalable Aggregate Network Attestation}.
\newblock In {\em ACM CCS'16}.

\bibitem{armknecht2013security}
F.~Armknecht~et al.
\newblock {A security framework for the analysis and design of software
  attestation}.
\newblock In {\em ACM CCS'13}.

\bibitem{asokan2015seda}
N.~Asokan~et al.
\newblock {SEDA: Scalable Embedded Device Attestation}.
\newblock In {\em ACM CCS'15}.

\bibitem{becher2006tampering}
A.~Becher~et al.
\newblock {Tampering with motes: Real-world physical attacks on wireless sensor
  networks}.
\newblock In {\em SPC'06}.

\bibitem{bernstein2006curve25519}
D.~J. Bernstein.
\newblock {Curve25519: new Diffie-Hellman speed records}.
\newblock In {\em PKC'06}.

\bibitem{bilel2012hybrid}
B.~R. Bilel~et al.
\newblock Hybrid cpu-gpu distributed framework for large scale mobile networks
  simulation.
\newblock In {\em IEEE/ACM DS-RT'12}.

\bibitem{brasser2015tytan}
F.~Brasser~et al.
\newblock {TyTAN: Tiny trust anchor for tiny devices}.
\newblock In {\em DAC'15}.

\bibitem{burchfield2007maximizing}
T.~R. Burchfield~et al.
\newblock Maximizing throughput in zigbee wireless networks through analysis,
  simulations and implementations.
\newblock In {\em LOCALGOS'07}.

\bibitem{castelluccia09:homomorphicAggregation}
C.~Castelluccia~et al.
\newblock {Efficient and provably secure aggregation of encrypted data in
  wireless sensor networks}.
\newblock In {\em TOSN'09}.

\bibitem{conti2009mobility}
M.~Conti~et al.
\newblock Mobility and cooperation to thwart node capture attacks in manets.
\newblock In {\em JWCN'09}.

\bibitem{conti2010smallville}
M.~Conti~et al.
\newblock The smallville effect: social ties make mobile networks more secure
  against node capture attack.
\newblock In {\em ACM MSWiM'10}.

\bibitem{di2010securing}
R.~Di~Pietro~et al.
\newblock {Securing Mobile Unattended WSNs against a Mobile Adversary}.
\newblock In {\em IEEE SRDS'10}.

\bibitem{dull2015high}
M.~D{\"u}ll~et al.
\newblock {High-speed Curve25519 on 8-bit, 16-bit, and 32-bit
  microcontrollers}.
\newblock {\em Designs, Codes and Cryptography}, 2015.

\bibitem{eldefrawy2012smart}
K.~Eldefrawy~et al.
\newblock {SMART: Secure and Minimal Architecture for (Establishing Dynamic)
  Root of Trust}.
\newblock In {\em NDSS'12}.

\bibitem{francillon2014minimalist}
A.~Francillon~et al.
\newblock A minimalist approach to remote attestation.
\newblock In {\em DATE'14}.

\bibitem{ho2010distributed}
J.-W. Ho.
\newblock {\em Distributed detection of node capture attacks in wireless sensor
  networks}.
\newblock INTECH, 2010.

\bibitem{hu03:secureAggregationWSN}
L.~Hu~et al.
\newblock {Secure Aggregation for Wireless Networks}.
\newblock In {\em SAINT'03}.

\bibitem{ibrahimdarpa}
A.~Ibrahim~et al.
\newblock {DARPA: Device Attestation Resilient to Physical Attacks}.
\newblock In {\em ACM WiSec'16}.

\bibitem{katz2008aggregate}
J.~Katz and A.~Y. Lindell.
\newblock {Aggregate Message Authentication Codes}.
\newblock In {\em CT-RSA'08}.

\bibitem{kohnhauser2016secure}
F.~Kohnh{\"a}user and S.~Katzenbeisser.
\newblock {Secure Code Updates for Mesh Networked Commodity Low-End Embedded
  Devices}.
\newblock In {\em ESORICS'16}.

\bibitem{kovah2012new}
X.~Kovah~et al.
\newblock {New results for timing-based attestation}.
\newblock In {\em IEEE S\&P'12}.

\bibitem{li2011viper}
Y.~Li~et al.
\newblock {VIPER: verifying the integrity of PERipherals' firmware}.
\newblock In {\em ACM CCS'11}.

\bibitem{morgner2016all}
P.~Morgner~et al.
\newblock {All Your Bulbs Are Belong to Us: Investigating the Current State of
  Security in Connected Lighting Systems}.
\newblock {\em arXiv'16}.

\bibitem{noorman2013sancus}
J.~Noorman~et al.
\newblock {Sancus: Low-cost Trustworthy Extensible Networked Devices with a
  Zero-software Trusted Computing Base}.
\newblock In {\em USENIXSec'13}.

\bibitem{park2012smatt}
H.~Park~et al.
\newblock {SMATT: Smart Meter ATTestation Using Multiple Target Selection and
  Copy-Proof Memory}.
\newblock In {\em FTRA CSA'12}.

\bibitem{park2016ain}
Y.~Park~et al.
\newblock {This Ain't Your Dose: Sensor Spoofing Attack on Medical Infusion
  Pump}.
\newblock In {\em WOOT'16}.

\bibitem{przydatek03:sia}
B.~Przydatek~et al.
\newblock {SIA:} secure information aggregation in sensor networks.
\newblock In {\em ACM SenSys'03}.

\bibitem{sailer2004design}
R.~Sailer~et al.
\newblock {Design and Implementation of a TCG-based Integrity Measurement
  Architecture}.
\newblock In {\em USENIXSec'04}.

\bibitem{skorobogatov2012physical}
S.~Skorobogatov.
\newblock Physical attacks and tamper resistance.
\newblock In {\em Introduction to Hardware Security and Trust}. Springer, 2012.

\bibitem{slawomir2016gattacking}
J.~Slawomir.
\newblock {GATTacking Bluetooth Smart devices}.
\newblock In {\em Black Hat USA}, 2016.

\bibitem{stmrtc}
{STMicroelectronics}.
\newblock {AN3371 Application note -- Using the hardware real-time clock (RTC)
  in STM32 F0, F2, F3, F4 and L1 series of MCUs}, 2012.

\bibitem{taban2008efficient}
G.~Taban and V.~Gligor.
\newblock Efficient handling of adversary attacks in aggregation applications.
\newblock In {\em ESORICS'08}.

\bibitem{mspusersguide}
{Texas Instruments}.
\newblock {MSP430x5xx and MSP430x6xx Family -- User's Guide Chapter 24.2.4 RTC
  Protection.}

\bibitem{wenger20138}
E.~Wenger, T.~Unterluggauer, and M.~Werner.
\newblock 8/16/32 shades of elliptic curve cryptography on embedded processors.
\newblock In {\em INDOCRYPT'13}.

\end{thebibliography}

\begin{appendix}

\section{Protocol Extensions}

\subsection{Leader Election Protocol}
\label{appendix:leaderelection}

\begin{figure*}[b!]
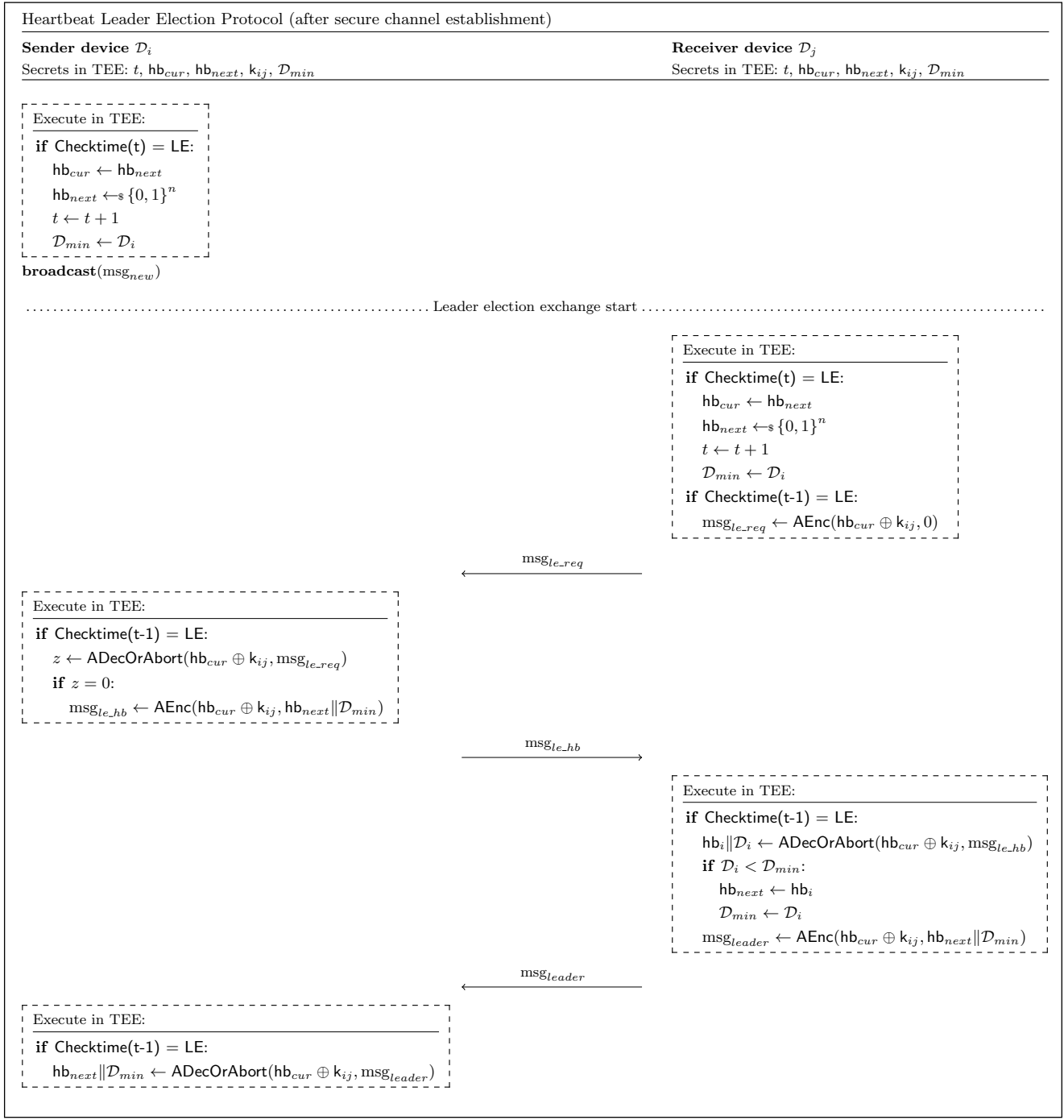

\begin{center}
\fbox{%
\scalebox{0.85}{
\procedure[]{Heartbeat Leader Election Protocol (after secure channel establishment)}{%
\textbf{Sender device $\device{i}$}      \<                    \< \textbf{Receiver device $\device{j}$} \\
\text{Secrets in TEE: $\now$, $\curHB$, $\nextHB$, $\key_{ij}$, \device{min}}\<     \< \text{Secrets in TEE: $\now$, $\curHB$, $\nextHB$, $\key_{ij}$,
\device{min}} \\[][\hline]\\
%
%
  \dbox{\begin{subprocedure}\procedure{\TEE}{
  \textrm{\pcif \timecheckLE{t}:}\\
  \t \text{$\curHB \leftarrow \nextHB$}\\
  \t \text{$\nextHB \sample \bin^n$} \\
  \t \text{$\now \leftarrow \now + 1$} \\
  \t \text{$\device{min} \leftarrow \device{i}$}
}
\end{subprocedure}}\\
\text{{\bf broadcast}($\msg_{new}$)} \pclb
\pcintertext[dotted]{Leader election exchange start}
\<\< \dbox{\begin{subprocedure}\procedure{\TEE}{
     \textrm{\pcif \timecheckLE{t}:}\\
     \t \text{$\curHB \leftarrow \nextHB$}\\
     \t \text{$\nextHB \sample \bin^n$} \\
     \t \text{$\now \leftarrow \now + 1$} \\
     \t \text{$\device{min} \leftarrow \device{i}$}\\
     \textrm{\pcif \timecheckLE{t-1}:}\\
     \t \text{$\msg_{le\_req} \leftarrow \AEnc{\curHB \xor \key_{ij}}{0}$ } 
}
\end{subprocedure}}\\
%
%
%
                    \<  \sendmessageleft*{\text{$\msg_{le\_req}$}} \\
\dbox{\begin{subprocedure}\procedure{\TEE}{
  \textrm{\pcif \timecheckLE{t-1}:}\\
  \t  \text{$z \leftarrow \ADecAbort{\curHB \xor \key_{ij}}{\msg_{le\_req}}$} \\
  \t \text{\pcif $z =0$:}\\
  \t \t \text{$\msg_{le\_hb} \leftarrow \AEnc{\curHB \xor \key_{ij}}{\nextHB \| \device{min}} $}
}
\end{subprocedure}}\\
                    \< \sendmessageright*{\text{$\msg_{le\_hb}$}} \\ 
\<\< \dbox{\begin{subprocedure}\procedure{\TEE}{
     \textrm{\pcif \timecheckLE{t-1}:}\\
     \t \text{$\hb_{i}\|\device{i} \leftarrow \ADecAbort{\curHB \xor \key_{ij}}{\msg_{le\_hb}}$}\\
     \t \textrm{\pcif $\device{i} < \device{min}$:}\\
     \t \t \text{$\nextHB \leftarrow \hb_{i}$}\\
     \t \t \text{$\device{min} \leftarrow \device{i}$}\\
     \t  \text{$\msg_{leader} \leftarrow \AEnc{\curHB \xor \key_{ij}}{\nextHB \| \device{min}} $}
}
\end{subprocedure}} \\
                    \< \sendmessageleft*{\text{$\msg_{leader}$}} \\ 
\dbox{\begin{subprocedure}\procedure{\TEE}{
  \textrm{\pcif \timecheckLE{t-1}:}\\
  \t  \text{$\nextHB \| \device{min} \leftarrow \ADecAbort{\curHB \xor \key_{ij}}{\msg_{leader}}$} 
}
\end{subprocedure}}\\
}}
}
%
%
\end{center}
\caption{Leader election protocol.\label{protocol:leader}}
\end{figure*}

\para{Heartbeat Transmission Extension}
The leader election protocol is shown in Figure~\ref{protocol:leader} and extends the heartbeat transmission phase described in \S~\ref{sec:protocol:heartbeat}. We henceforth divide each time period $T_1, T_2, T_3, ...$ of length $\Tslot$ in two phases: the heartbeat phase, whose length is $\Tslot_{hb}$ (formerly $\Tslot$), and the leader election phase, whose length is $\Tslot_{le} = \Tslot - \Tslot_{hb}$. Furthermore, we assume that the function $\timecheck(t)$ returns the constant $\HB$ if $T_t \leq T_{clock} < T_t + \Tslot_{hb}$, the constant $\LE$ if $T_t + \Tslot_{hb} \leq T_{clock} < T_{t+1}$, and $\false$ otherwise.

Every device $\device{i}$ that did not receive a heartbeat within $\Tslot_{hb}$, indicated by $\timecheckLE{t}$, will generate its own heartbeat $\nextHB^{i}$, set the current leader device id to its own id ($\device{min}^{i} \leftarrow \device{i}$), and update its time pointer by one. In a next step $\device{i}$ informs its neighbors about the new heartbeat with a message $\msg_{le}$.


Two devices that already initialized the leader election phase negotiate the heartbeat as follows.
First, a leader election request message $\msg_{le\_req}$ is generated by
$\device{j}$ that contains a session key to secure the remaining communication.
Then, $\device{i}$ sends the smallest received device id $\device{min}^{i}$ and the
corresponding heartbeat to $\device{j}$. Initially these are $\device{i}$'s own id and generated heartbeat.
Device $\device{j}$ will then compare its previous smallest id $\device{min}^{j}$
with the just received id.
If $\device{min}^{i}$ < $\device{min}^{j}$, $\device{j}$ will update $\device{min}^{j}$  to $\device{min}^{i}$
and set $\nextHB^{j}$ to  $\nextHB^{i}$.
Finally, $\device{j}$ will inform $\device{i}$ of the result of the comparison,
which is also stored by $\device{i}$.
Both devices will then continue to further broadcast
the new heartbeat. The protocol terminates 
implicitly, once the smallest device id has been identified.
We note that a leader, who is absent
during the heartbeat phase, can rejoin by participating in the leader election phase.
In \S~\ref{evaluation:dynamic}, we analyze the effectiveness of the 
leader election protocol.

\para{Security}
The leader election protocol uses the same two-key mechanism, i.e., the session key constructed by heartbeat and channel key, as the original heartbeat protocol to secure all messages. This makes it impossible for an adversary \Adv to synthesize or to decrypt a message that is accepted or sent by healthy devices. Otherwise, \Adv could break the IND-CTXT or IND-CPA security of the encryption scheme. Hence, the actual leader election process can only be hindered, yet not controlled by \Adv.

\subsection{Attestation of Software Integrity}
\label{appendix:softwareintegrity}
To achieve a secure attestation of hardware and software the following extensions
to \ourname are required:

\para{Enrollment Phase Extension}
In the enrollment phase, the network operator selects an arbitrary software integrity measurement function \textsf{Measure()} and stores its implementation in the \ac{TEE} of each device $\device{1}, ..., \device{n}$ in the network. Traditionally, these mechanisms measure the integrity of a software by computing a hash value over its binary code~\cite{sailer2004design}, though recent approaches are also able to measure the runtime behavior of a software, for the purpose of detecting sophisticated code-reuse attacks~\cite{abera2016c}. In the following, we abstract from these implementation details and use \textsf{Measure()} as a black box that takes an input $i$, e.g., a description of what to measure, and generates a measurement $m$, which represents the current software state of a device (\device{i}: \textsf{Measure($i$) $\rightarrow m$}).



\para{Attestation Protocol Extension}
Before invoking the attestation protocol, $\mathcal{O}$ specifies a set of trustworthy software states $tss$. $T\hspace{-0.3mm}ss$ consists of multiple (input, measurement)-pairs ($tss = \{(i_1, m_1), (i_2, m_2), ..., (i_x,m_x)\}$) and a description which network device should use which input (e.g., devices from type $1$ should use $i_1$, etc.). A pair $(i_k, m_k)$ in $tss$ indicates that the expected measurement for the input $i_k$ is $m_k$ (\textsf{Measure($i_k$) = $m_k$}). In this way, $tss$ specifies all measurements that are permitted by $\mathcal{O}$, e.g., because they represent the correct and most recent software states.

During the execution of the attestation protocol, $tss$ is distributed to all devices in the network. For this purpose, $\mathcal{O}$ initially incorporates $tss$ into $\msg_{\verifier}$ ($\msg_{\verifier}$ $\leftarrow$ $\AEnc{\devicekey_i}{ts\|n\|tss}$). In a similar way, by incorporating $tss$ into $\msg_{att}$ ($\msg_{att} \leftarrow \AEnc{\key_{cur}}{ts\|tss}$), devices forward $tss$ to neighboring devices. Afterwards, each device \device{i} measures its local software configuration by extracting its appropriate $(i_k, m_k)$ pair and executing the measurement function \textsf{Measure()} with the input $i_k$ in its \ac{TEE}. Subsequently, \device{i} checks whether the output generated by \textsf{Measure()} matches $m_k$ and if this is the case continues with the execution of the attestation protocol, as explained in Section~\ref{sec:protocol:attestation}. If both values do not match, \device{i} invokes a recovery routine, which allows the device to restore to a trustworthy state by performing a secure code update protocol with $\mathcal{O}$. Executing this extended attestation protocol, $\mathcal{O}$ receives a $msg_{res}$ that only contains ids of devices that are in a trustworthy software and uncompromised hardware state. Note that the protocol could easily be further extended to precisely report devices which are in an untrustworthy software but uncompromised hardware state, e.g., by introducing an additional $\msg_{agg\_s}$ and $\msg_{res\_s}$ that is specifically generated and aggregated by untrustworthy devices and transmitted to $\mathcal{O}$.



\para{Security}
The security of the protocol extension results from the security of the main protocol (\S~\ref{sec:protocol:security}), the secure hardware properties (\S~\ref{preliminaries}), and the adversary model (\S~\ref{preliminaries}). Since the protocol extension is executed in the \ac{TEE} of devices, malware is unable to tamper with the protocol code, execution, or any stored protocol data (e.g., secret keys). Thus, an adversary, who compromises the software of a device, is only able to prevent protocol execution or manipulate the input/output to/from the protocol. However, preventing protocol execution has no influence, since untrustworthy devices stop executing the attestation protocol, anyway. Manipulating the input or output to or from the protocol has no affect, as all inputs and outputs are secured using authenticated encryption with secrets that are only accessible within the \ac{TEE}. Additionally, all inputs and outputs are dependent on a session-specific timestamp $ts$ issued by $\mathcal{O}$. Therefore, replay attacks are likewise worthless. These measures also prevent Dolev-Yao network adversaries from compromising security. By contrast, a physical attacker is able to tamper with the protocol code, data, or execution. However, as explained in the security analysis of the main protocol (\S~\ref{sec:protocol:security}), a physical attacker is unable to obtain the current heartbeat \curHB, which is required to participate in the attestation protocol or heartbeat protocol.

\subsection{Efficient Attestation Report Aggregation}
\label{appendix:aggregation}

\para{Security} As already shown in the security analysis of the
aggregation protocol (\S~\ref{sec:protocol:security}), \Adv is unable
to exchange any message with healthy devices during attestation.
This argument also holds for the efficient aggregation scheme, as only
the aggregation inside the \ac{TEE} is modified and not the protocol itself.
Consequently, the security of the efficient aggregation scheme, depends on the
hardness of attestation report itself.
An attestation report is accepted
if at least $n / 2$ valid attests are contained in the report. By assumption
$\Adv$ is only allowed to compromise up to $c < n/2 - s$ devices and thus, can only
compute up to $c$ valid attests. The remaining $n/2 - c$ attests have to be
guessed by $\Adv$. The security of our aggregation scheme is formalized in
Theorem~\ref{theorem:saggregation}.

\label{theorem:saggregation}
\begin{theorem}
Assuming collision resistance of the hash function, any PPT $\Adv$,
compromising up to ${c < n/2 - s}$ devices can successfully forge
an efficient attestation report that is accepted by $\mathcal{O}$ with
probability of at most $2^{-s}$ for any ${n>2\cdot s}$.
\end{theorem}

\para{Proof Sketch} The attestation report consists of two bit vectors,
the first vector annotates the
devices included in the network and the second vector annotates the actual
attests (each attest is a single bit in the attest vector).
To successfully include one additional attest into the report,
$\Adv$ has to set an additional bits in the device vector and to guess
the correct bit in the attest vector. A single mismatch between the 
aggregate computed by $\mathcal{O}$ and the reported aggregate
results in a reject of the attestation report.
As the position of an attest bit for a single device is computed by
$\textsf{compress}(\hash(\devicekey_i\|ts))$, we
observe that $\Adv$ could break the collision resistance 
of the hash function, if $\Adv$ would achieve non-negligible advantage
in guessing an attest bit correctly without access to the device key. 
Assuming a uniform distribution of the attestation bit, 
$\Adv$ will guess its position correctly with probability $\frac{1}{n_s}$.
However, $\Adv$ can follow a better strategy than randomly guessing all positions
of the $n/2-c$ bits that are required for a valid attestation report.
We note that due to the relatively small set of bit positions, collisions
between multiple devices are likely. The best strategy the \Adv can follow
is thus, to guess collisions with the $c$ bits that \Adv can set correctly 
in the attest vector.
A collision with any of the attest bits occurs with probability of at most
$\frac{c}{n_s}$ (collisions within the attests of compromised devices are
also possible).
With this strategy, $\Adv$ can achieve a winning probability of
at most $(\frac{c}{n_s})^{n/2-c}$. We observe that
$\frac{c}{n_s} = \frac{n/2-s}{n+s} \leq \frac{1}{2}$
and by assumption $n/2 - c \geq s$ and thus, 
$\Adv$ wins the game with probability of less than $2^{-s}$.

We remark that for the sake of technical simplicity of the proof,
the attestation vector is set to a fixed length $n_s = n + s$.
This is required to make it a hard task for $\Adv$ to guess the zero bits
in the attest vector for smaller $n$, when setting all bits in the device
vector. For lager $n$, $n_s$ could be chosen smaller than $n+s$.




\end{appendix}

\end{document}